\documentclass[twocolumn, times, trackchanges]{aastex63}
\usepackage{amsmath}

\usepackage{booktabs}
\usepackage{courier}

\usepackage{subfigure}
\usepackage{soul}

\usepackage{csquotes}

\received{}
\revised{}
\accepted{}
\submitjournal{}

\shorttitle{Evolution of Clumpy SFGs}
\shortauthors{Sok et al.}
\graphicspath{{./}{figures/}}

\begin{document}

\title{Finite Resolution Deconvolution of Multi-Wavelength Imaging of 20,000 Galaxies in the COSMOS Field:\\ The Evolution of Clumpy Galaxies Over Cosmic Time}

\correspondingauthor{Visal Sok}
\email{sokvisal@yorku.ca}

\author[0000-0003-0780-9526]{Visal Sok}
\affiliation{Department of Physics and Astronomy, York University, Toronto, ON M3J 1P3, Canada}

\author[0000-0002-9330-9108]{Adam Muzzin}
\affiliation{Department of Physics and Astronomy, York University, Toronto, ON M3J 1P3, Canada}

\author[0000-0002-9655-1063]{Pascale Jablonka}
\affiliation{Laboratoire d’Astrophysique, Ecole Polytechnique F\'ed\'erale de Lausanne (EPFL), Observatoire, CH-1290 Versoix, Switzerland}

\author[0000-0002-7248-1566]{Z. Cemile Marsan}
\affiliation{Department of Physics and Astronomy, York University, Toronto, ON M3J 1P3, Canada}

\author[0000-0002-3503-8899]{Vivian Y. Y. Tan}
\affiliation{Department of Physics and Astronomy, York University, Toronto, ON M3J 1P3, Canada}

\author[0000-0002-2250-8687]{Leo Alcorn}
\affiliation{Department of Physics and Astronomy, York University, Toronto, ON M3J 1P3, Canada}

\author[0000-0001-9002-3502]{Danilo Marchesini}
\affiliation{Department of Physics and Astronomy, Tufts University, Medford, MA 06520, USA}

\author[0000-0001-7768-5309]{Mauro Stefanon}
\affiliation{Leiden Observatory, Leiden University, NL-2300 RA Leiden, Netherlands}

\nocollaboration{9}



\begin{abstract}

Compact star-forming clumps observed in distant galaxies are often suggested to play a crucial role in galaxy assembly. In this paper, we use a novel approach of applying finite resolution deconvolution on ground-based images of the COSMOS field to resolve 20,185 star-forming galaxies (SFG) at $0.5<z<2$ to an angular resolution of 0.3", and study their clumpy fractions. A comparison between the deconvolved and \textit{HST} images across four different filters shows good agreement and validates the deconvolution. We model spectral energy distributions using the deconvolved 14-band images to provide resolved surface brightness and stellar mass density maps for these galaxies. We find that the fraction of clumpy galaxies decreases with increasing stellar masses, and with increasing redshift: from ${\sim}30\%$ at $z\sim0.7$ to ${\sim}50\%$ at $z\sim1.7$. Using abundance matching, we also trace the progenitors for galaxies at $z\sim0.7$ and measure the fractional mass contribution of clumps toward their total mass budget. Clumps are observed to have a higher fractional mass contribution toward galaxies at higher redshift: increasing from ${\sim}1\%$ at $z\sim0.7$ to ${\sim}5\%$ at $z\sim1.7$. Finally, the majority of clumpy SFGs have higher specific star formation rates (sSFR) compared to the average SFGs at fixed stellar mass. We discuss the implication of this result to \textit{in-situ} clump formation due to disk instability. 


\end{abstract}



\section{Introduction} \label{sec:intro}

Over the last 3 decades, we have developed a broad understanding of the stellar mass build-up within galaxies over much of cosmic time. The star formation rate density is known to decrease over the past ten billion years (e.g., \citealt{Lilly:1996, Madau:1996, Madau2014}), while the bulk of the global stellar mass density was formed during the epoch of peak cosmic star formation at $1<z<3$ (e.g., \citealt{Dickinson:2003,Muzzin:2013b, McLeod:2021}). However, a complete picture of where and how stars are formed within galaxies remains elusive as it is still unclear which mechanisms are driving the stellar mass assembly. 

When looking at galaxy populations across cosmic time, we find that there is a transition in galaxy morphologies toward higher redshifts. In particular, peculiar galaxies that host kiloparsec-scale and clump-like structures outnumber Hubble-Sequence galaxies at $z>1$ \citep{Elmegreen2007}. These clumpy structures are star-forming regions as indicated by their enhanced specific star formation rates (sSFR) (e.g., \citealt{Wuyts2012, Hemmati2014}), and their detection in high-resolution observations of H$\alpha$ emission line maps (e.g., \citealt{Wuyts2013, Mieda2016}). While there is no formal definition, compact star-forming structures in distant galaxies are generally dubbed as \enquote{clumps}, and their host galaxies as \enquote{clumpy galaxies}. Clumpy star-forming galaxies (SFGs) have been studied for both field and lensed galaxies, with a variety of multi-wavelength data, including the rest-frame UV and optical imaging, emission line maps, and CO observations (e.g., \citealt{Elmegreen2005, Elmegreen2007, Elmegreen2009a, Elmegreen2009b, Forster2011, Wuyts2012, Murata2014, Livermore2015, Guo2015, Shibuya2016, Cava2018, Zanella2019}). Despite their ubiquity at high-$z$, the origin and evolution of clumps are not well constrained by observational studies due to the limiting resolution of our observing facilities. Size estimates of star-forming clumps can vary from tens of parsecs to a few kiloparsecs based on observations at $z\sim1{-}4$ (e.g., \citealt{Livermore2015}; \citealt{Soto2017}; \citealt{Cava2018}). At these scales, even the \textit{Hubble Space Telescope} (\textit{HST}) can only marginally resolve these structures at its limiting angular resolution of ${\sim}0.1"$, or a physical resolution of ${\sim}1$ kpc, at $z\sim2$.

As a result, the formation and evolution of clumps can be explained within a wide range of theoretical models. The mechanisms for clump formation are associated with both \textit{in-situ} and \textit{ex-situ} origins. Such clumpy and irregular morphologies can be the result of \textit{ex-situ} processes such as mergers, which is supported by some observational studies \citep{Guo2015, Calabro:2019}. It is also not clear whether clumps are just remnants of accreted satellites that have not been tidally disrupted (e.g.,  \citealt{Mandelker2014}; \citealt{Zanella2019}). The other line of evidence for their formation is that these clumps are not only observed in interacting systems, but also galaxies that exhibit a dynamic structure consistent with that of a rotating disk as identified on the basis of their kinematics (e.g., \citealt{Forster2009,Genzel2011,ForsterSchreiber:2018}). A majority of these high-redshift ($z\sim1{-}3$) disks are often observed to be gaseous and turbulent, with gas mass fraction of 20\%-80\% (e.g., \citealt{Daddi2010}; \citealt{Tacconi2010}) and intrinsic gas velocity dispersion of ${\sim}20{-}90$ km s$^{-1}$  (e.g., \citealt{Forster2009,Wisnioski2015,ForsterSchreiber:2018}). These two properties can lead to \textit{in-situ} processes where gravitational instabilities create large star-forming clumps: large gas dispersion results in large Jeans lengths, with the high gas fraction sustaining on-going star formation (e.g., \citealt{Toomre:1964, Dekel_2009}). 

Due to their unique mode of star formation during the main epoch of massive galaxy building, clumps are often suggested to have a central role in galaxy assembly. However, the details of how clumps affect the growth of galaxies is poorly understood at the moment. For example, the fate of clumps is inconclusive; clumps can be long-lived stellar structures that migrate inward and coalesce onto the progenitor of present-day galactic bulges (e.g., \citealt{2016ASSL..418..355B}, \citealt{Mandelker2014, Mandelker2017}), or short-lived phenomena, where strong stellar feedback disperse the gas and unbind the stellar system on time-scale of ${<}100$ Myr \citep{Hopkins:2012}. 

In order to decipher the nature of star-forming clumps at high-$z$, high resolution and multi-wavelength data are needed to resolve these structures and to infer their physical properties, respectively. At present, \textit{HST} is the only observatory capable of providing such data. While \textit{HST} has now been in service for approximately three decades, there are only a few existing fields with truly deep and multi-wavelength imaging in more than a handful of filters (e.g., GOODS; \citealt{Giavalisco:2004}). Data from the GOODS-South and ERS \citep{Windhorst:2011} have already been exploited for resolved stellar populations studies \citep{Wuyts2012}. While recent data such as the Hubble Frontier Field (HFF; \citealt{Lotz:2017}) are available, the amount of future data that will be comparable to these remains small. On the other hand, larger cameras on ground-based observatories can take advantage of the large field of view, which allow for a larger volume of the sky to be surveyed. However, ground-based observatories are subjected to atmospheric turbulence, and the seeing of these images are effectively degraded, rendering them of limited use for resolved studies of galaxies. 

Image deconvolution has often been explored as a solution to reverse the atmospheric and instrumental blurring effect. One challenge with deconvolution is that a sampled image cannot be fully deconvolved without violating the sampling theorem \citep{Magain:1998}. A solution is to deconvolve images to a finite resolution, such that the resolution of the deconvolved images is compatible with the new sampling interval. Further pioneering work by \cite{2016A&A...589A..81C} onto this deconvolution technique has shown that it is possible to achieve close to space-based resolutions in deconvolved ground-based images (see \citealt{Cantale:2016}). The new algorithm, finite resolution deconvolution (FIREDEC), has two main features that are advantageous compared to previous deconvolution algorithms. (1) FIREDEC deconvolves to a finite resolution, ensuring that solutions are well-sampled and constrained, and (2) FIREDEC introduces a new regularization scheme to characterize noise, allowing the algorithm to better distinguish signal from noise.

In this paper, we present FIREDEC as an alternative method to obtain high resolution images of distant galaxies. We use image deconvolution to provide resolved stellar properties of high-$z$ ($0.5<z<2$) galaxies within the Cosmic Evolution Survey (COSMOS; \citealt{Scoville:2007}) field. The original ground-based images have a typically seeing of $0.6{-}0.8$ arcsec, and are deconvolved to a target resolution of 0.3". This resolution corresponds to a physical size of ${\sim}2.4$ kpc at our redshift range. COSMOS is the ideal field to perform this study as it covers an area of ${\sim}2~\mathrm{deg}^2$, and has arguably the most impressive array of multi-wavelength coverage (30+ photometric bands). 


The paper is structured as follows. Section \ref{sec:data} describes the observational datasets and our selection of high-$z$ star-forming galaxies. In Section \ref{sec:decon}, we give a brief introduction to image deconvolution, and provide a comparison between deconvolved and \textit{HST} images. In Section \ref{sec:method_maps}, we discuss our methodology for constructing spatially-resolved stellar population maps, while Section \ref{sec:method_identify} discusses how the surface mass density and surface brightness maps are used to identify and classify clumpy galaxies. We present the measured fraction of clumpy galaxies in relation to redshift and global properties of galaxies such as stellar masses and star formation rates in Section \ref{sec:clumpfrac}. In Section \ref{sec:abundmatch}, we trace the progenitors of clumpy SFGs using abundance matching, and present the fractional mass contribution of star-forming clumps as galaxies evolve. Finally, we discuss the correlation between star formation rates and clumpy morphologies, and its implications on the origin of star-forming clumps. We adopt the following cosmological parameters $\Omega_M=0.3$, $\Omega_\Lambda=0.7$, and $H_{\mathrm{o}}=70$ km s$^{-1}$ Mpc$^{-1}$. All magnitudes are quoted in the AB magnitude system.

\section{Data Description} \label{sec:data}

We make use of the COSMOS/UltraVISTA catalog from \cite{Muzzin2013}. The catalog was obtained using PSF-matched photometry in 30 photometric bands that span from GALEX NUV to Spitzer MIPS 24 $\mu$m. A total of 262,615 sources were detected within the UltraVISTA $K_s$-band imaging, which reaches a depth of $K_{s,\mathrm{tot}} < 23.7$. The field covers a total area of 1.6 deg$^2$. A detailed description of the photometric catalog construction, along with measurements of the photometric redshift, stellar mass, and UV+IR-determined star formation rate, is presented in \cite{Muzzin2013}. 
 
\subsection{Selection of the Photometric Data} \label{sec:datasel}

The photometric data from the COSMOS/UltraVISTA catalog consisted of 30 filters with varying seeing. For example, the UV and NIR data have seeing that typical ranges between $0.5{-}1.2$" \citep{Muzzin2013}. Ideally, we would want to deconvolve all the available imaging of COSMOS. However, deconvolving to a higher angular resolution is the equivalent of recovering high frequency components. The deconvolution of data with both poor and good image qualities (IQ) to a similar (higher) resolution introduces signal-to-noise ratio (S/N) variations, as deconvolution introduces high frequency components within its solution. It can therefore become difficult to reconcile structures within deconvolved images when using a non-homogeneous data that range in image qualities, and can bias our measurements of stellar properties from SED modeling.


\begin{table}[!t]
    \renewcommand{\arraystretch}{1.15}
    \begin{tabular}{ccc}
    \toprule
    Filter & Seeing (") & Reference  \\
    (1) & (2) & (3)  \\
    \midrule
    $B_j$   & 0.71 - 0.78 & \cite{Capak2007}  \\
    $V_j$   & 0.74 - 0.84 & - \\
    $r^+$   & 0.78 - 0.85 & - \\
    $z^+$   & 0.81 - 0.91 & - \\
    IA484   & 0.54 - 0.72 & - \\
    IA527  & 0.59 - 0.67 & - \\
    IA624   & 0.70 - 0.81 & - \\
    IA738   & 0.72 - 0.80 & - \\
    IB427   & 0.66 - 0.75 & - \\
    IB505   & 0.71 - 0.84 & - \\
    $Y$       & 0.82 - 0.86 & \cite{McCracken2012} \\
    $J$       & 0.81 - 0.85 & - \\
    $H$       & 0.78 - 0.82 & - \\
    $K_s$   & 0.77 - 0.82 & - \\
    \bottomrule
    \end{tabular}
    \caption{Summary of the photometric data used within this study. As summarized in Section \ref{sec:datasel}, only images that have uniform and good image qualities are selected. The seeing variations are obtained from \cite{Muzzin2013}.}
    \label{tab:seeinginfo}
\end{table}

A solution is to deconvolve photometric data that has good and relatively homogeneous image qualities. Within the large number of filters available in COSMOS, we choose not to use the $g^+$ filter of the Subaru/SuprimeCam as the data have the worst seeing compared to the others (1.01" to 1.20"). We also omit the $i^+$-band image as its longer integration time caused most stars to be saturated and unusable for point spread function modeling. Finally, most of the medium-band photometric data have poor image qualities (e.g., median IQ $\gtrsim0.9$), and are therefore omitted. 

The total number of filters that are used within this study is 14. These consist of four optical broadband images ($B_jV_jr^+z^+$) and the six IA and IB optical intermediate-band images from the Subaru/SuprimeCam. The four near-infrared images, $YJHK_s$, from the UltraVISTA survey are also included. Our selected data have a total seeing range of $0.5{-}0.9"$, as summarized in Table \ref{tab:seeinginfo}. 



\subsection{Selection of Galaxy Sample}

The COSMOS/UltraVISTA catalog contained 262,615 sources. The parent galaxy sample is obtained using the flag, \enquote{\textit{use=1}}. Essentially, the \enquote{\textit{use=1}} flag excludes objects that are flagged as stars based on a color-color cut, objects that are badly contaminated by nearby bright stars, and objects that contain bad pixel regions. The total number of galaxies in the parent sample is 160,070.

Using the derived redshift, stellar mass, and star formation rate from the COSMOS/UltraVISTA catalog, we further filter suitable galaxies for the study. Specifically, we select star-forming galaxies at $0.5<z<2$, as this paper is focused on the evolution of clumpy star-forming galaxies. Also, only galaxies that satisfy a target S/N cut are included in the study. The choice for the target S/N is discussed in Section \ref{sec:masscomp} as well as the criteria needed to ensure that deconvolution can recover new spatially-resolved information within the galaxies. 

\subsubsection{Limiting S/N for Image Deconvolution} {\label{sec:masscomp}}

Image deconvolution is an iterative process where the original image is resampled and pixel intensities are redistributed until the solution minimizes a defined cost function (see Section \ref{sub:firedec}). As a result, it is of limited value to deconvolve objects that are already faint and have low S/N, as deconvolution cannot recover fine scale structures that are not already at high S/N in the original images. 


The recovered S/N per pixel by deconvolution depends on the surface brightness of the galaxy. In the ideal situation, we would know the shape and size of galaxies \textit{a priori} and can select galaxies based on its surface brightness to ensure that we achieve some minimum S/N per pixel. In practice, such morphological parameters are not known from ground-based images. Therefore, we use the integrated S/N of the galaxies in the $K_s$-band as a proxy for what their surface brightness may be. The integrated S/N is calculated as the ratio between the total flux and error as determined by SExtractor's flux\_auto. For galaxies between $0.5<z<2$, the $K_s$-band is the best single-filter proxy for stellar mass as it probes the rest-frame NIR for galaxies at $z\sim0.5$ and the rest-frame optical at $z\sim2$. Our tests shows that the amount of information recovered from deconvolved galaxies with S/N $<$ 20 is limited, so deconvolution is only applied to galaxies with S/N $> 20$ in the $K_s$-band. 

\begin{figure}[t!]
\centering
\includegraphics[width=0.9\columnwidth]{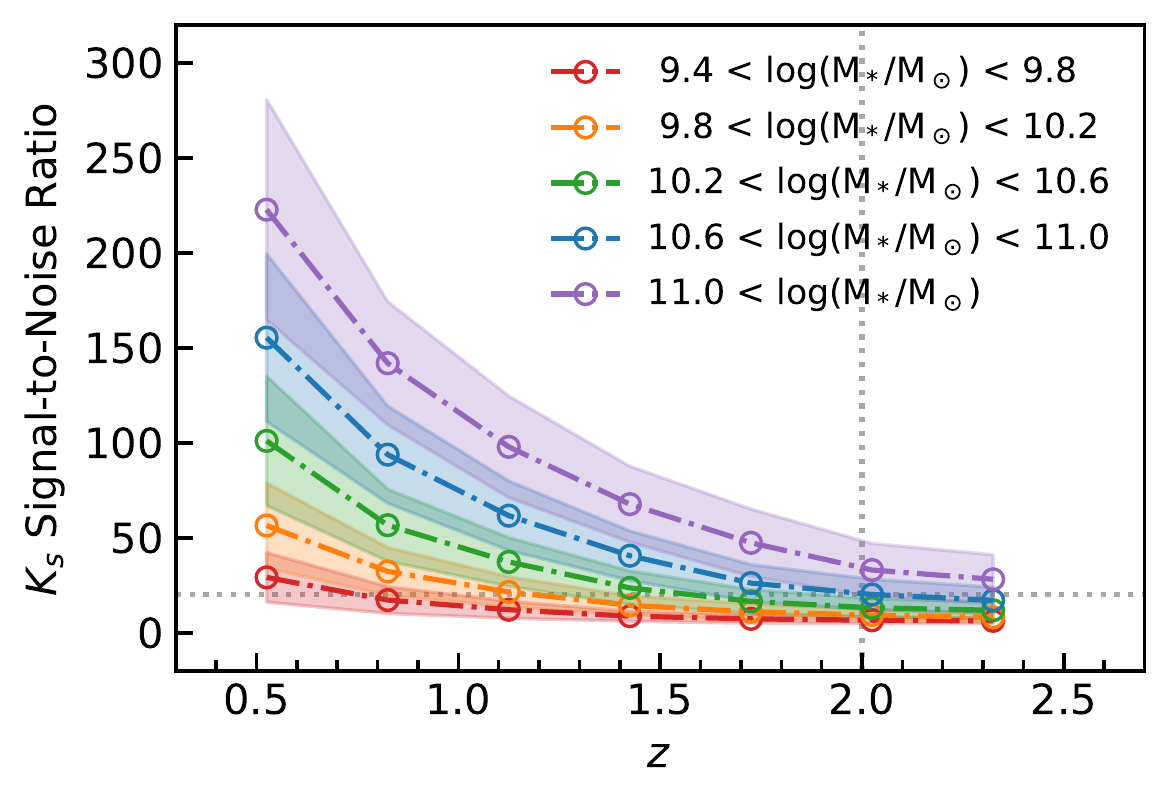}
 \caption{ The $K_s$ signal-to-noise ratio as a function of redshift. Galaxies are binned into different redshift and stellar mass bins. Only galaxies between the redshift of $0.5<z<2$ and galaxies with S/N ($K_s$) greater than 20 are selected, as denoted by the dotted lines.  }
 \label{fig:limitingsnr}
\includegraphics[width=0.975\columnwidth]{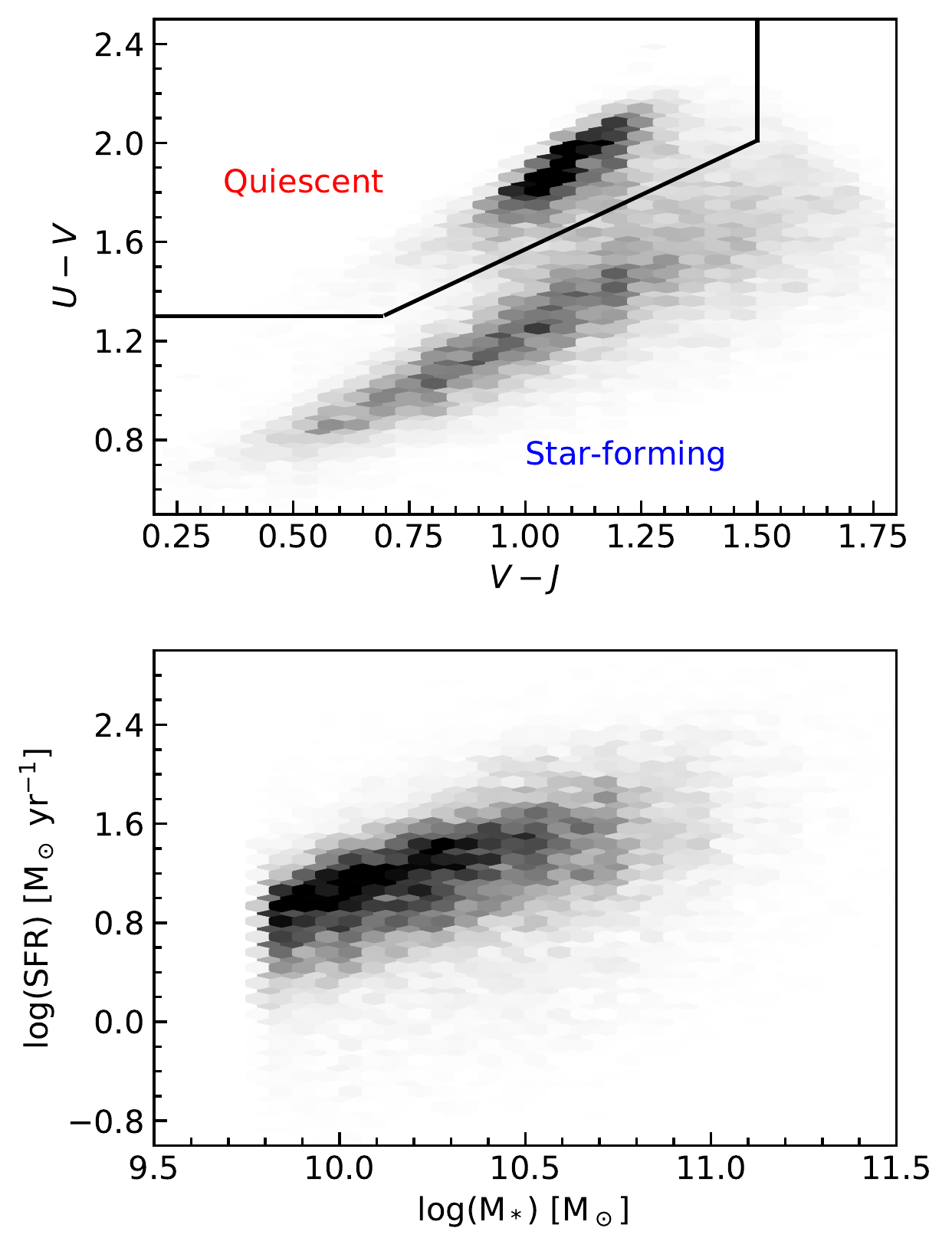}
 \caption{The UVJ diagram and the star-forming main sequence. Top: The distribution of all the galaxies in COSMOS at $0.5<z<2$ within the UVJ diagrams excluding those with $\log$(M$_*$/M$_\odot) < 9.8$ and $K_s$-band S/N $< 20$. Bottom: The distribution of the selected star-forming galaxies in the SFR-M plane.}
 \label{fig:uvjdiag}
\end{figure}

Figure \ref{fig:limitingsnr} shows the distribution of galaxies based on the integrated $K_s$-band S/N and redshift for different mass bins. The horizontal dotted line shows the S/N cut we use, and galaxies below the dotted line are excluded in the analysis in the remainder of the paper. We also omit galaxies with $\log$(M$_*$/M$_\odot) < 9.8$ as the majority of them have an integrated S/N below the limiting S/N. Similarly, we exclude all galaxies above $z > 2$ as only massive galaxies with $\log$(M$_*$/M$_\odot) > 11$ satisfy the required S/N. The total number of galaxies from the parent sample that satisfies the S/N, mass and redshift cuts is 32,903. 

Note that our sample is a S/N limited sample, not a mass-limited sample. Using the integrated S/N as a proxy assumes that all galaxies at a given magnitude have similar light profiles, which is not necessarily true. This is however unavoidable as the recovered S/N across the deconvolved images will depend on the surface brightness of the galaxy. In order to obtain a uniform, mass-complete sample with redshift, one would need to model how the surface brightness depends on mass and redshift. However, based on our selection criteria, only low-mass galaxies with low surface brightness are excluded at higher redshifts, and this should cause only minimal bias in the results.


\subsubsection{Defining Star-Forming Galaxies}
The main focus of the study is on the stellar populations of galaxies during their lifetime prior to quenching, therefore the sample is further separated into star-forming and quiescent galaxies. The morphological bi-modality of galaxy populations has been well studied in the local universe and is observed to persist to higher redshifts (e.g., \citealt{Muzzin:2013b}). Several methods have already been developed to distinguish the two populations. In particular, the classification based on the $U-V$ vs. $V-J$ (UVJ) diagram has been used in many previous studies (e.g.,  \citealt{Williams2009, Whitaker2011, Muzzin2013}). The separation of star-forming and quiescent galaxies based on the UVJ diagram has been shown to correlate with the separation based on UV+IR-determined specific star formation rates (hereafter sSFR$~\equiv~$SFR/M$_*$ , e.g. \citealt{Williams2009}) and based on SED-determined sSFRs (e.g. \citealt{Williams2010}). The inverse of the sSFR defines a timescale for the formation of the stellar population within the galaxy, which is why it is often used as a diagnostic of quiescence. Galaxies are labeled as star-forming if their rest-frame $(U-V)$ and $(V-J)$ colors satisfy,

\begin{equation}
\begin{split}
(U-V)&\leq 1.3 \\
(U-V)&\leq 0.88(V-J)+0.69 \\
(V-J)&\geq 1.6
\end{split}
\end{equation}

In the top panel of Figure \ref{fig:uvjdiag}, we plot the distribution of galaxies satisfying the limiting S/N and redshift criteria onto the UVJ diagram. The bi-modality of galaxy populations can be clearly seen in the UVJ diagram, where quiescent galaxies lie along the upper envelope (above the solid black line) and star-forming galaxies along the lower envelope. In total, the number of deconvolved star-forming galaxies that satisfy the color, redshift, and S/N criteria is 22,156. In the bottom panel, we show the distribution of these galaxies in the SFR-M$_*$ plane, where the SFRs are determined from the UV and IR fluxes (details of how these are derived are presented in \citealt{Muzzin2013}). The total number of SFGs in this study is approximately 1-2 orders of magnitude larger in comparison to previous \textit{HST} studies of the clumpy fraction of SFGs at similar redshift ranges (e.g., ${\sim}700$ in \citealt{Wuyts2012} and ${\sim}3300$ in \citealt{Guo2015}). 

\section{Image Deconvolution} \label{sec:decon}

\subsection{Finite Resolution Deconvolution}\label{sub:firedec}

One of the main features of our deconvolution algorithm is that images are deconvolved to a finite resolution. Deconvolved images therefore have a narrower point spread function (PSF) of their own. Such finite resolution deconvolution ensures that solutions are well-constrained and well-sampled. The following equations are generalized from \cite{2016A&A...589A..81C}. We encourage the reader to review \cite{2016A&A...589A..81C} for the technical discussion on finite resolution deconvolution.


Here, we use an asterisk to denote the convolution operator. An image ($D$) taken from a ground-based optical telescope can be mathematically expressed as, 
\begin{equation}
    D = M*PSF + Z, 
\end{equation}
where $PSF$ is the point spread function and $Z$ is the noise. $M$ is a model of the image that is unaffected by atmospheric and instrumental blurring. The PSF of an image can further be expressed as, 
\begin{equation}\label{eqn:fires}
    PSF = g*P,
\end{equation}
where \textit{g} is the target PSF, taken to be a 2D Gaussian function, and \textit{P} is the kernel that transforms \textit{g} to \textit{PSF}. Within FIREDEC, the construction of $P$ consists of two fits: an analytic fit and then a numerical fit. The analytic fit models a Moffat-like profile to obtain the main characteristics of the PSF (i.e., the central core and the \enquote{wings} of PSF). Instead of fitting an exact Moffat profile, the analytic function is estimated as a sum of three elliptical Gaussians. The numerical fit is then needed to further model structures of the PSF that are not properly captured by the analytic fit. The numerical fit is obtained by deconvolving the residual of the analytic fit by the target PSF. 

When images are sampled onto a finite grid of pixels and affected by noise, deconvolution can become an ill-posed problem where many solutions are compatible with the data. However, a solution to deconvolution can be obtained by minimizing a cost function that include a regularization term, 
\begin{equation}
    C(M) = \sum_i \Big[\frac{D - (PSF*M)}{\sigma}\Big]^2_i + \lambda H, 
    \label{eqn:cfunc}
\end{equation}
where the summation goes over each pixel element and $\sigma$ represents the noise map for the observed image. Since FIREDEC deconvolves to a finite resolution, the cost function is written as, 
\begin{equation}
    C(B) = \sum_i \Big[\frac{D - (P*B)}{\sigma}\Big]^2_i + \lambda H, 
    \label{eqn:cfunc_firedec}
\end{equation}
where $B = M*g$. For both equations, the first term is a chi-squared term and the second term is a regularization term, $H$, weighted by a constant Lagrange parameter, $\lambda$. The purpose of the second term is to penalize high frequency components that arise during deconvolution. Within FIREDEC, the regularization term is defined to be, 

\begin{equation}
    H = \sum_i\Big( \frac{B - \Tilde{B}}{\sqrt{1+B}} \Big)_i^2, 
\end{equation} 
where $\Tilde{B}$ is the denoised version of $B$, or $\Tilde{B} = \phi(B)$. The denoising function implemented in FIREDEC, $\phi$, is based on wavelet transform and is reviewed by \cite{2016A&A...589A..81C}. Their implementation of the regularization term takes into accounts the dynamical range of astronomical images. As the S/N within an image can vary strongly, such implementation ensures that high S/N regions are treated in a similar way as low S/N regions during the deconvolution process, without having to define a spatially varying $\lambda$. 

The determination of the Lagrange parameter is also crucial, as it ultimately dictates the importance of the regularization term within Equation \ref{eqn:cfunc_firedec}. Traditionally, finding the optimal value for $\lambda$ requires testing different values until a relative good deconvolution is achieved. This is checked by visually inspecting the noise within the deconvolved images. A quantitative analysis is used within this study to determine $\lambda$, and is outlined in Appendix \ref{append:lagrange}. Note that two different Lagrange parameters are determined in Appendix \ref{append:lagrange}. A Lagrange parameter is needed for the numerical fit of the PSF, where the fit is obtained by deconvolving the residuals of the analytic fit with the target PSF using a cost function similar to Equation \ref{eqn:cfunc_firedec}. A separate Lagrange parameter is required for the deconvolution of the ground-based image ($D$) to obtain a model of the image with the target PSF of $g$ ($B$).

\begin{figure}[t!]
\centering
\includegraphics[width=0.975\columnwidth]{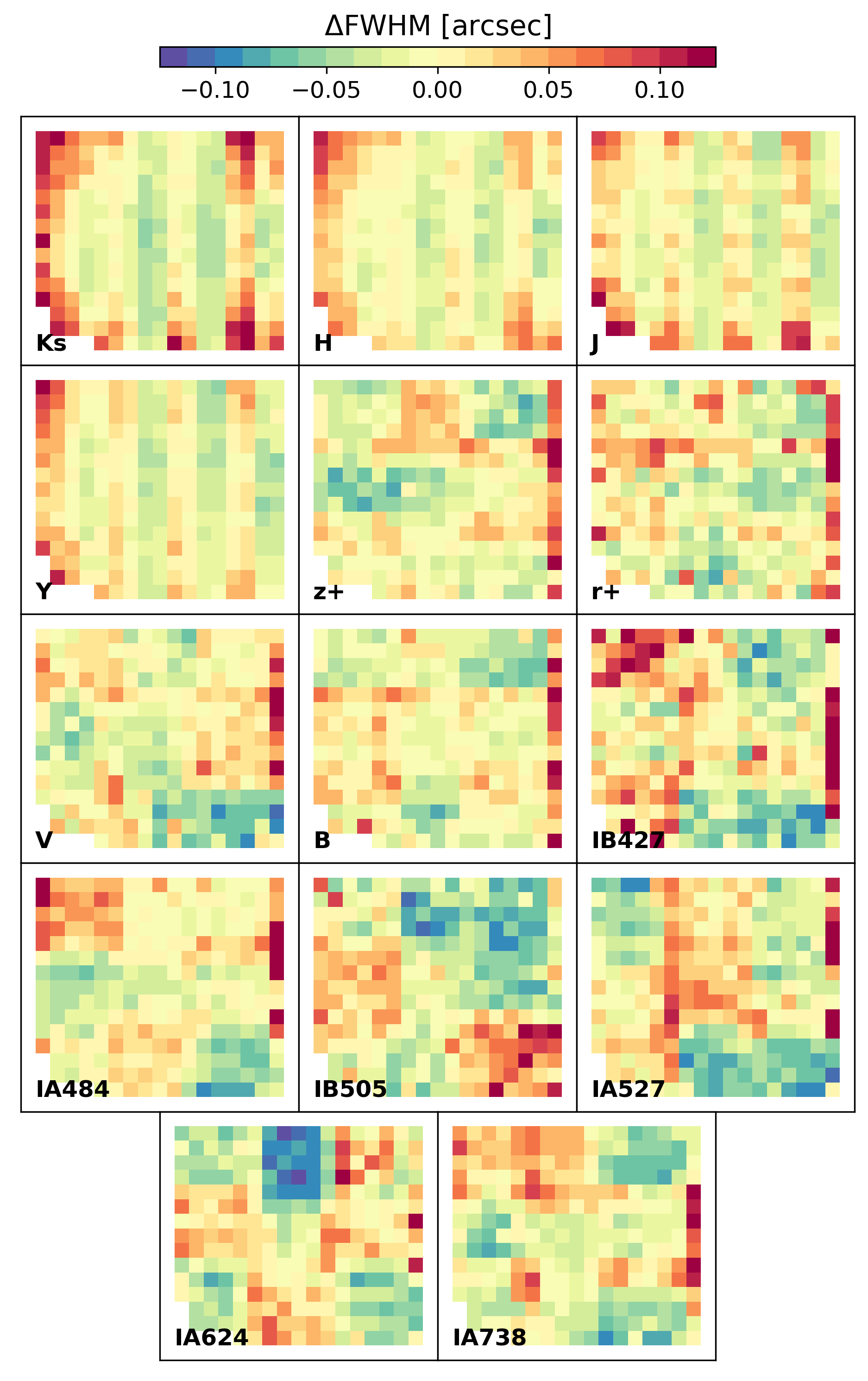}
 \caption{The seeing variations of COSMOS for eaphotometric filter. We bin each mosaic into pixel regions of $25$ arcmin$^2$, and use SExtractor to calculate the FWHM for all the stars. The colormap shows the differences between the median FWHM within each pixel and the median FWHM over the mosaic.  }
 \label{fig:psfvariation}
\end{figure}

\subsection{Deconvolution in Practice}

Instead of deconvolving the entire image of the COSMOS field in a given filter (which can become computationally intensive), we instead deconvolve smaller image cutouts that are centered on each galaxy. Doing so also ensures that that we account for any variations in the PSF across the full COSMOS field. Indeed, observations of the COSMOS field consisted of multiple observing runs that extended over several years, leading to varying image qualities. 

Figure \ref{fig:psfvariation} shows the seeing variations for each filter. Each map covers an approximate area of $1.4~\mathrm{deg} \times 1.2~\mathrm{deg}$, i.e., the full COSMOS field. The maps are binned into $17\times15$ different regions, where each region is $25$ arcmin$^2$. The colormap shows the difference between the median FWHM within a region and the median FWHM of the field (i.e., $\Delta$FWHM). Note that in the optical Subaru filters, the full mosaic consists of a $3 \times 3$ grid of SuprimeCam pointings. The variation in image qualities in each of those pointings relative to each other is quite visible in some filters (e.g., IB505 and IA624). There is also PSF variations within the pointings (which is common in large mosaic cameras like SuprimeCam), demonstrating the importance of determining a local PSF when performing deconvolution.

Different deconvolution kernels are created based on the local PSF in order to deconvolve images to the same angular resolution. SExtractor is used to find all potential PSF stars within the field. The brightest stars that are flagged as unblended and unsaturated are selected as suitable stars for PSF modeling. We also verify that each star does not have any close companions by inspecting the residual of its PSF fit for structures that are not associated with the star. It should be noted that the deconvolution kernel for each galaxy is modeled using the closest suitable PSF star to the galaxy. The same star may not be necessarily used as the PSF star in different filters as its magnitude can vary.

At $0.5<z<2$, the effective radius of galaxies are observed to be ${\sim}3{-}5$ kpc (e.g., \citealt{Wel2014}), corresponding to an angular size of ${\sim}1{-}2"$. We use an image cutout size of 7.8" by 7.8", which is roughly 4 effective radii bigger than the estimated size of galaxies. The target resolution is set as 0.3", which corresponds to a gain of a factor of $2{-}3$ in resolution per band. A resolution of 0.3" also enables us to probe a physical size of ${\sim}2.4$ kpc at $z\sim1$. While this is approximately several times bigger than the estimated size of clumps (${\sim}1$ kpc; e.g., \citealt{Soto2017}), Section \ref{sec:clumpfrac} shows that star-forming clumps can still be probed, even at this coarser resolution. 

The target resolution of 0.3" is chosen to avoid oversampling pixels and mitigate noise that arise from deconvolution. Our ground-based images typically have a FWHM of ${\sim}0.7"$ and a pixel scale of 0.15", corresponding to a sampling factor of around ${\sim}4$ for the PSF. Deconvolving to a higher resolution would require for smaller sampling interval by each pixel to allow for higher frequency components, which include both signal and noise. At our target resolution, each pixel is only resampled into $2^2 = 4$ sub-pixels. The $4{\times}$ resampling factor allows for the angular resolution to be improved to a level that enable us to partially resolve galaxies and detect clumpy structures, but also does not oversample the data to a point where it is difficult to distinguish clumps and noise. As we demonstrate in the remainder of the paper, our deconvolved images look like \textit{HST} in similar filters and our measurements are also consistent with \textit{HST}-based studies. This shows that this choice in resolution is optimal for maintaining sufficient S/N per sub-pixel, yet still recovering key features of galaxy properties that are required to classify them as clumpy or not.


\subsection{Examples of Deconvolved Images}\label{sec:deconv_eg}

We compare the deconvolved images to \textit{HST} images in order to see how well the deconvolution parameters are determined for the COSMOS data. Note that a direct comparison to validate whether the deconvolved images match the ground-based images (given the modeled PSF) was also presented in \cite{Cantale:2016} at lower $z$ ($\leq 0.8)$. 

The COSMOS field was imaged in the F814W filter, with partial imaging in the F475W filter. There are also available F606W images from the CANDELS/COSMOS survey \citep{Grogin2011, Koekemoer2011}, and F160W images from COSMOS-DASH survey \citep{Mowla2019}. In total, there is an area of ${\sim}0.03~\mathrm{deg}^2$ that has all 4 filters. We emphasize that \textit{HST} images are only used to verify structures that are resolved by FIREDEC. These images are not used as additional constraints for deconvolution. 

\subsubsection{Quantitative Comparison with F814W}

\begin{figure}[t!]
\centering
\includegraphics[width=\columnwidth]{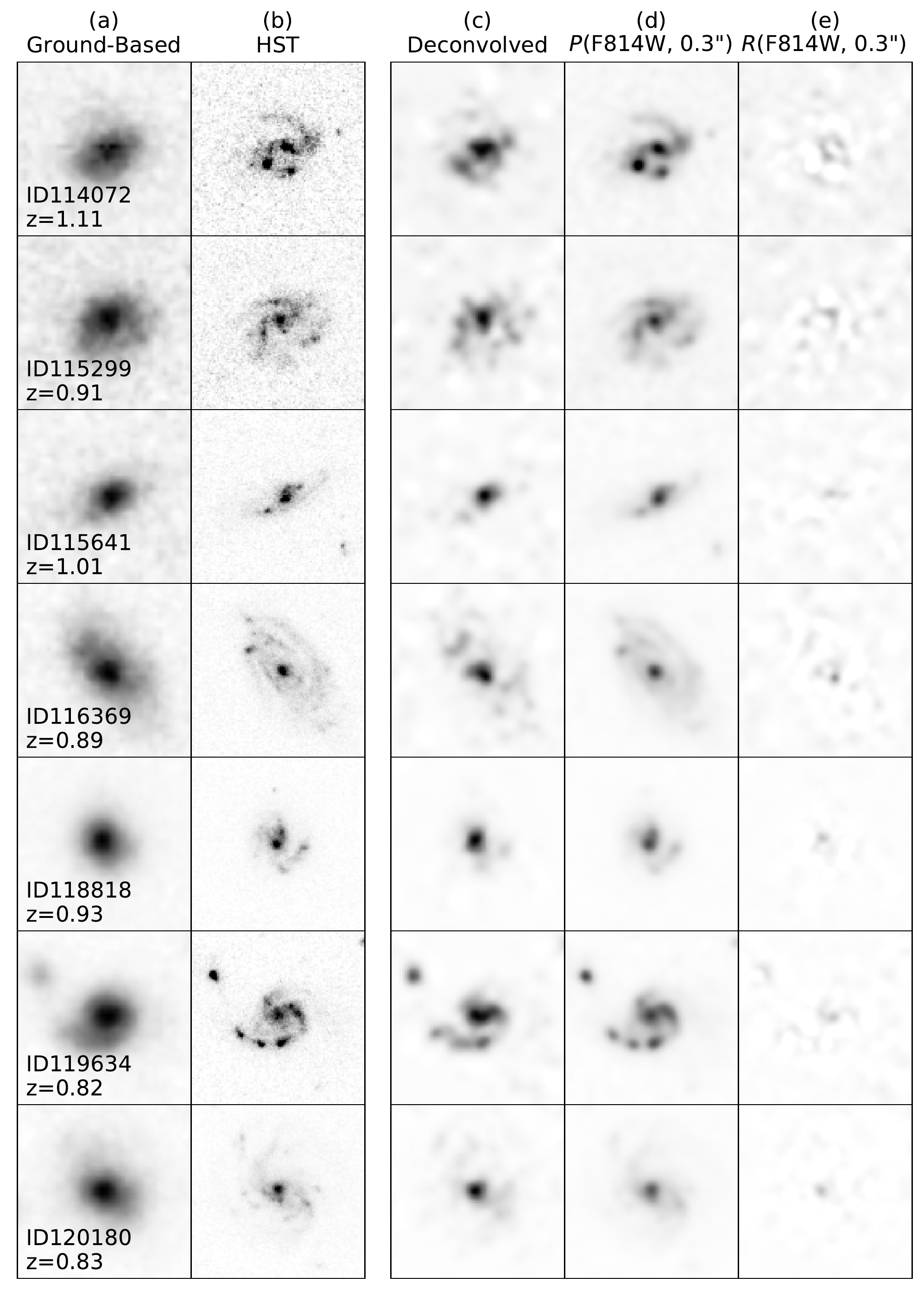}
\caption{Comparison between the deconvolved $z^+$ and ACS/\textit{HST} F814W images. Columns (a) and (b) show the ground-based and \textit{HST} images, while the deconvolved images are shown in column (c). We PSF-match the \textit{HST} images to the deconvolved images at an angular resolution of 0.3" (column d). The differences between the deconvolved and PSF-matched \textit{HST} images are shown as $R(\mathrm{F814W}, 0.3")$ in column (e). In general, the residual maps illustrate that deconvolved structures are consistent with \textit{HST} images, with only minor structures within the central regions.} 
 \label{fig:quantResi}
\end{figure}

A quantitative comparison between the Subaru $z^+$ and F814W images is complicated by the fact that the two filters are not truly identical. The wavelength coverage of F814W extends from ${\sim}7000{-}9500$ \r{A} and its transmission curve peaks around 7500 \r{A}, while the Subaru broadband $z^+$ filter spans from ${\sim}8000{-}10000$ \r{A}. The PSF of both images also differ: $z^+$-band images are affected by atmospheric blurring and their PSF can be characterized as a Moffat profile, while the PSF of F814W images is diffraction-limited, with large and extended PSF wings. Some differences between the deconvolved $z^+$ images and their \textit{HST} counterparts can therefore be attributed to the effects of the morphological K-correction, where galaxy morphologies change as a function of wavelength, and differing PSFs. Keeping these points in mind, we present the quantitative comparison between the deconvolved and \textit{HST} images. 

\begin{figure*}[t!]
\centering
\includegraphics[width=0.95\textwidth]{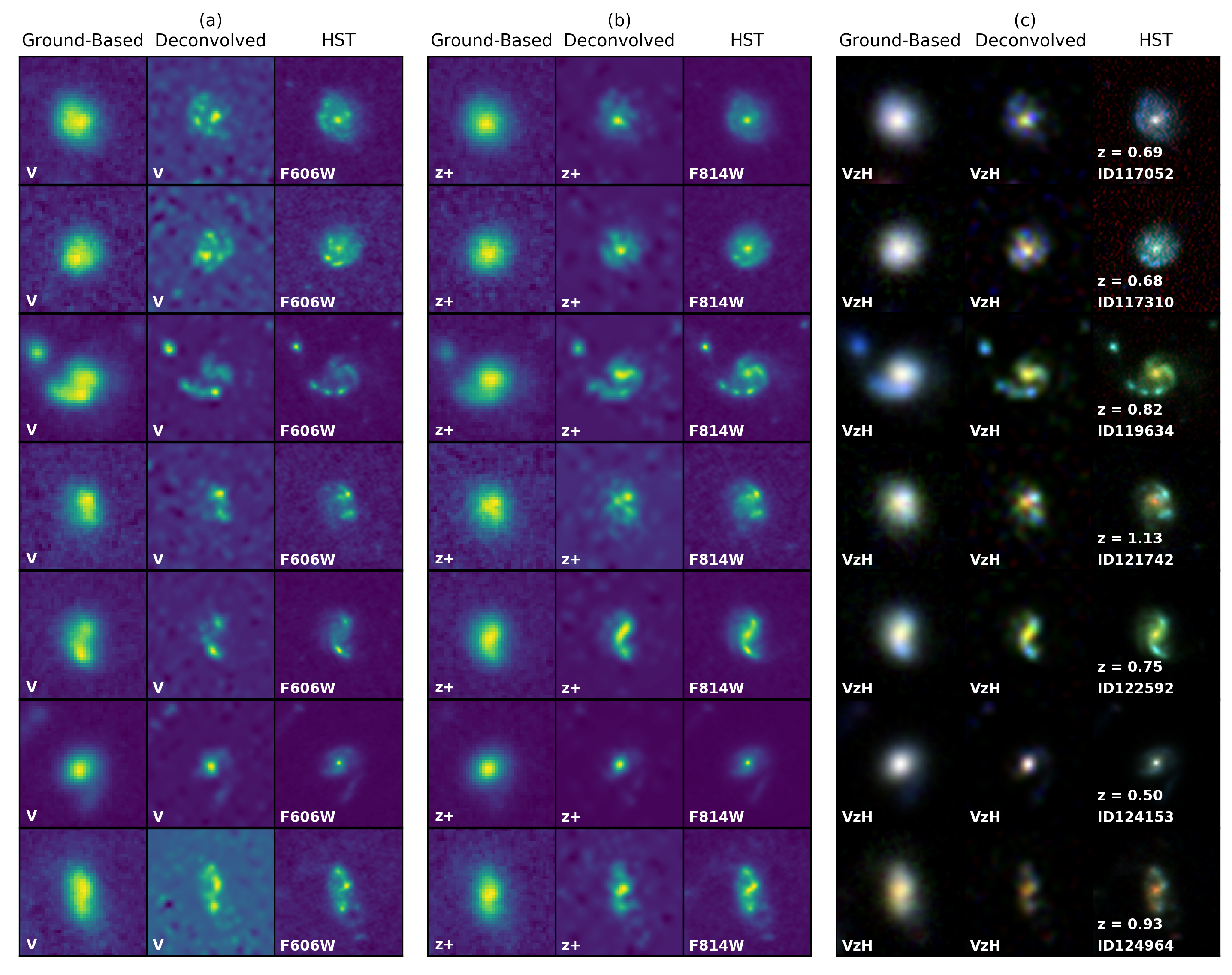}
 \caption{Examples of deconvolution using multi-wavelength inages. Each row corresponds to a different galaxy. Panels (a) and (b) show examples of deconvolved images at a resolution of 0.3" in the rest-frame UV and optical, respectively, for galaxies between $0.5<z<1$. Panel (c) shows the composite images. Within each panel, we show the ground-based, deconvolved and \textit{HST} image for that galaxies from the left to right column. These galaxies are chosen mainly for their clumpy morphologies as observed in the deconvolved images. The \textit{HST}/ACS F606W and F814W images are PSF-matched to the F160W image, with an angular resolution of 0.18". Similarly, the \textit{HST} composite images are shown at a resolution of 0.18".  }
 \label{fig:multi-dec}
\end{figure*}
    
\begin{figure*}[t!]
\centering
\includegraphics[width=0.95\textwidth]{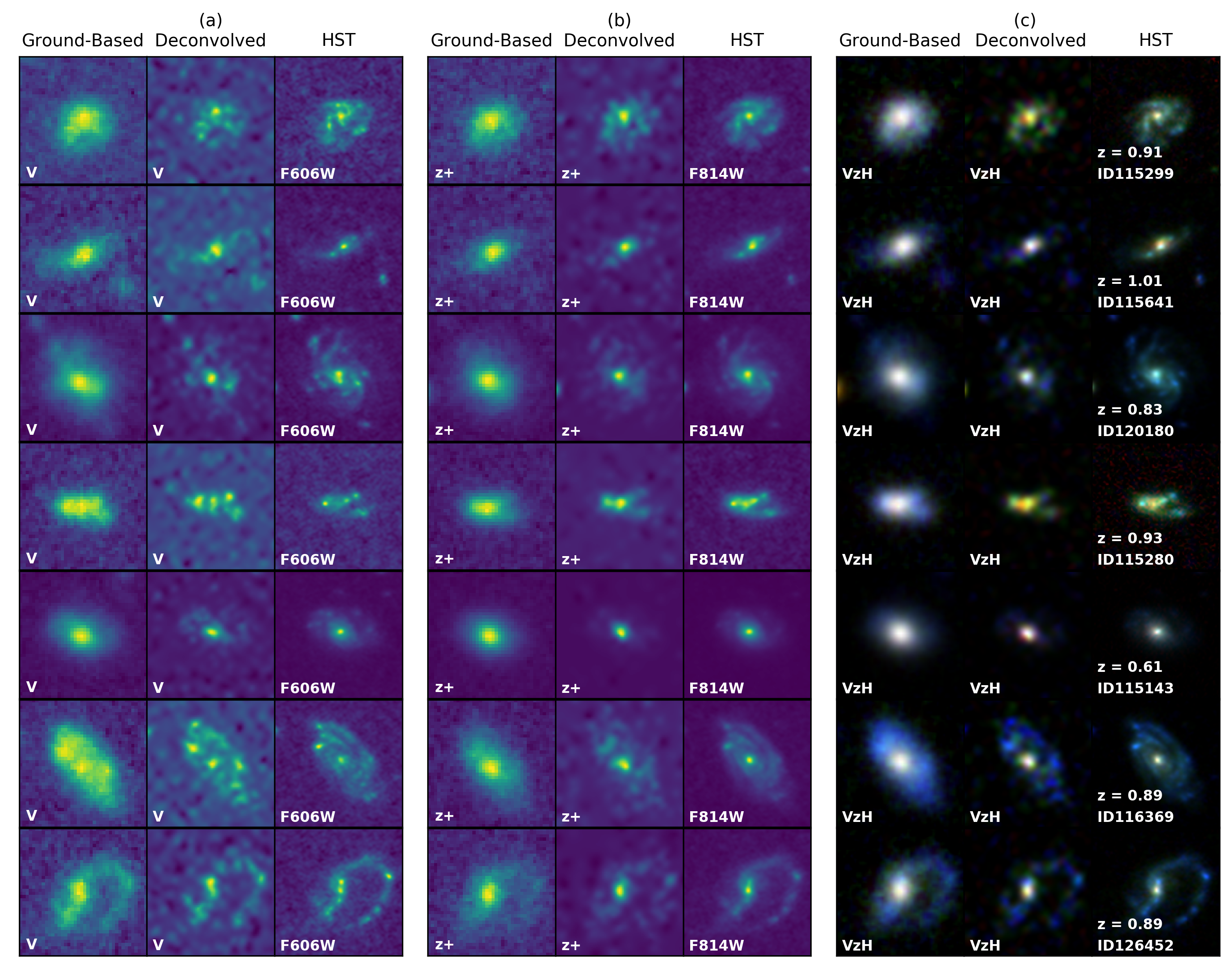}
 \caption{More examples of image deconvolution along with the comparison to the \textit{HST} images. The \textit{HST} images are at an angular resolution of 0.18". }
  \label{fig:multi-dec-2}
\end{figure*}

The following notations are used to simplify the descriptions: $P(x, a)$ represents PSF-matching an image $x$ to a resolution of $a$, while $R(x, a)$ represents taking the differences between an image $x$ and the deconvolved images at that resolution. The F814W images are first degraded to match the resolution of the deconvolved images. This is done by convolving them by a Gaussian function so that their resolution becomes 0.3". The PSF-matched \textit{HST} images are shown as $P(\mathrm{F814W}, 0.3")$ in column (d) of Figure \ref{fig:quantResi}, and the differences between them and the deconvolved images are shown in column (e) as $R(\mathrm{F814W}, 0.3")$. We also account for any misaligned astrometry by cross-correlating the \textit{HST} images with the deconvolved images. In general, we find that the residuals of $R(\mathrm{F814W}, 0.3")$ are fairly uniform near the outskirts of galaxies, with minor structures near the central regions. These residuals are most likely due to the imperfect characterization of the PSF since it is impossible to know the exact profile of the PSF at the location of each galaxy due to the PSF variations across the field. However, the outer regions of the galaxies within the residual maps show negligible structures and indicates that key aspects such as clumps are clearly captured by deconvolution. In particular, examples such as ID116369 and ID118818 further illustrate that clumps are barely visible in the ground-based, but show nicely within the deconvolved images.


\subsubsection{Multi-Wavelength Comparison}

Figures \ref{fig:multi-dec} and \ref{fig:multi-dec-2} compares the deconvolved multi-wavelength images of several galaxies with their corresponding \textit{HST} images. The information across the different filters are not used as additional constraints for deconvolution, but instead the images are deconvolved separately. Each row corresponds to a different galaxy, while the columns show the rest-frame UV and optical imaging, along with the composite image of each galaxy. We show galaxies that have been imaged in multiple filters by \textit{HST}. The galaxies are also selected mainly for their clumpy morphologies as observed in the deconvolved images. 

A visual inspection of the deconvolved and \textit{HST} images reveals that deconvolved images (FWHM = 0.3") are strikingly similar to their \textit{HST} counterparts (FWHM = 0.18"), even at the coarser resolution. Structures resolved in the deconvolved images are also observed within the \textit{HST} images. Examples such as ID119634, ID124963 and ID115641 demonstrate the potential of FIREDEC where deconvolution recovers distinct clumps that are barely discernible in the ground-based $V$ or $z^+$-band images. The deconvolved structures are also observed across different filters, which further indicates that these structures are real, and not artifacts of deconvolution. A closer inspection of the deconvolved images from Figures \ref{fig:multi-dec} and \ref{fig:multi-dec-2} also shows that deconvolution is limited by the S/N of the ground-based images. In cases where the ground-based images have low signal, the low contrast between signal and noise causes deconvolved noise to appear more prominently as clumpy structures. This is noticeable particularly within the background of low S/N images such as the deconvolved $V$-band of ID117052 and ID117310. We stress that this does not have an effect on our analyses as noise is not correlated across different filters. Combining multi-wavelength observations to model the spectral energy distributions mitigates the effects of noise in our study of star-forming clumps. Indeed, since clumps are identified using the rest-frame $U$ and $V$ surface brightness maps (see Section \ref{sec:method_identify}), noisy structures that are presented in single images are not detected within these maps. Finally, while individual clumps may not be resolved at the target resolution of 0.3", we iterate that the main goal of this paper is to broadly study the morphology of galaxies and to showcase the capability of finite resolution deconvolution. As Figures \ref{fig:multi-dec} and \ref{fig:multi-dec-2} show, deconvolution, while imperfect, offers a gain in angular resolution and presents a possible solution to obtain observational data that are similar in angular resolution to \textit{HST} imaging. 



\section{Methodology For Creating Spatially-Resolved Stellar Population Maps} \label{sec:method_maps}

The following sections give an overview of how multi-wavelength images are combined so that resolved stellar properties can be inferred from SED models. We explain how star-forming clumps are detected within the resolved stellar mass density and surface brightness maps, and give a formal definition for what constitutes as a clumpy star-forming galaxy.  

\subsection{Constructing Multi-Wavelength Deconvolved Maps}


\subsubsection{Segmentation Maps} \label{sec:segmap}
Segmentation maps are needed to distinguish different objects within an image cutout and assign specific pixels to them. To create these maps, we use a watershed segmentation algorithm from scikit-image \citep{Walt2014}. The watershed algorithm treats the intensity of a grey-scale image as an overturned topological surface. The algorithm then uses defined markers to flood the valley until different markers merge to create the boundary of the segmentation maps. In practice, the markers are defined to be the coordinate of objects within the image cutout. For each image cutout, we mask the background, and the masked image is then normalized by the negative of its sum to create the overturned topological map. The background of image is determined by 3-$\sigma$ clipping the image. 




\subsubsection{Misaligned Astrometry}
In general, COSMOS has good astrometry. The error associated with the astrometric solutions is small (e.g. $<0.2$" for the optical data and ~0.1" for the NIR data; \citealt{Capak2007, McCracken2012}). While this is sufficient for photometry, it is not ideal for resolved studies of galaxies. Notably, such offsets are roughly the size of clumps, and therefore can have significant impact on our clump analyses. This is particularly true for compact galaxies, where a misalignment at the shorter rest-frame UV wavelengths can result in the misidentification of off-center clumps.   

While the morphology of galaxies can change considerably between the optical and NIR, it is comparable between filters that are adjacent in wavelength. As a correction for potential offsets, we calculate the relative offsets in pixel coordinates by iteratively cross-correlating each photometric cutout to the adjacent photometric cutout at longer wavelength. This is applied on the ground-based images. To this end, we fix the astrometry for the $K_s$-band images, and align the other image cutouts with respect to it. We use the integrated S/N of the galaxy (taken from the UltraVISTA catalog) as an indicator of whether a correction should be made. If the integrated S/N for a given photometric cutout is less than 5, we do not apply any corrections as there is already very little signal within the image cutout.

\subsubsection{Pixel Binning} \label{sec:vorbin}
Modeling the spectral energy distributions on the pixel-by-pixel basis can lead to unreliable measurements of the underlying stellar populations as the photometry of individual pixels can have low S/N. A solution is to bin pixels, so that each bin has some minimum S/N and ensures that the correct stellar properties can be obtained. However, galaxies spatially exhibit a wide range of S/N, so binning methods such as the Voronoi binning technique \citep{Cappellari_2003} are used to homogenize the S/N across the galaxies, given a constraint on the minimum S/N. An example is shown in \cite{Wuyts2012}, where they binned pixels based on the \textit{H}-band in order to study the underlying stellar populations of galaxies at $z\sim2$. 

However, binning based on the NIR alone can cause localized signals such as clumps to spread over a large area as the morphology of galaxies can vary between the rest-frame UV to NIR. Recently, \cite{Fetherolf2020} improved the Voronoi binning technique by incorporating the S/N information across multiple filters in the binning process. Such binning technique takes into account the morphological changes, but requires high S/N data as additional constraints from multiple filters generally lead to bigger bin size (and subsequently lower angular resolution) for low S/N data. While the global S/N of an image does not change as flux is recentered during deconvolution, regions of low surface brightness that are not initially observed in ground-based images are uncovered by deconvolution. Since our galaxies are selected to have an integrated S/N as low as 20, we opt to bin pixels based on the photometric information from two filters, rather than multiple filters in order to optimize the S/N per bin and the bin size, while accounting for the morphological changes across wavelengths.

We modify \cite{Cappellari_2003}'s Voronoi binning technique so that there are two binning channels: one based on the filter that probes the rest-frame UV of the galaxy (i.e., dependent on the redshift) and the other based on the observed $K_s$-band. In the original Voronoi binning technique, a pixel is accreted to a bin if the addition of that pixel improves the S/N of the bin. The accretion of pixels stop once the bin reaches the desired S/N threshold. In our modified version, the binning process is similar, except a bin is decided if it reaches a minimum S/N in either one of the two bands. We define the minimum S/N to be 5. The pixel noise of each deconvolved filter is estimated by taking the standard deviation of the flux within empty apertures. These empty apertures are randomly placed across the deconvolved images, and have a diameter of 0.15", which correspond to the original pixel scale of the ground-based images. As alluded to earlier, binning also degrades the angular resolution of the data, so the inferred stellar population maps are rescaled using the deconvolved images in order to recover the deconvolved resolution of 0.3". This is discussed in Section \ref{subsec:rescale}. 



\subsection{Resolved Spectral Energy Distribution Modeling} \label{sec:seds}
We use EAZY \citep{Brammer2008} to compute the rest-frame $U$ and $V$ luminosities for each spatial bin for all galaxies. EAZY calculates the colors by integrating the best-fit SED through the redshifted filter curves over the appropriate wavelength range. We use the response curve defined in \cite{Maiz2006} for the $U$ and $V$ filters. We fix the redshift of each spatial bin to the photometric redshift of its respective galaxy. 

The stellar populations for each bin are determined by individually fitting \cite{Bruzual2003}'s stellar population synthesis models to the 14-band $B$-to-$K_s$ SED with FAST/C++\footnote{\href{https://github.com/cschreib/fastpp/releases/tag/v1.3.1}{https://github.com/cschreib/fastpp/releases/tag/v1.3.1}}. This is essentially the C++ version of FAST \citep{Kriek2009}. A \cite{Chabrier2003}'s initial mass function and \cite{Calzetti2000}'s dust extinction law are adopted, and we assume a uniform solar metallicity and exponentially declining star formation history. We allow $\log(\tau)$ to vary between 7.0 and 10.0, in increments of 0.2, and  $\log(t)$ to vary between 7.0 and 10.0, in increments of 0.1. We also restrict $t$ to be less than the age of the universe at the observed redshift of each galaxy. The visual attenuation $(A_V)$ of each fit is allowed to vary between 0 and 4.

Note that the SED models make no distinction between clumps and non-clumpy regions, and therefore assumes that clumps have the same properties as the surrounding region (e.g., the same star formation history; SFH hereafter). While this may not be necessarily true, little is known about the SFHs of high-$z$ clumps. We choose to use the $\tau$-model as it is commonly assumed for distant galaxies and it has been shown to recover stellar masses and SFRs with typical systematic offsets of ${\sim}0.3$ dex \citep{Lee:2018}.

\subsubsection{Recovering the Angular Resolution of 0.3"} \label{subsec:rescale}

Modeling the SEDs of each spatial bin allows us to construct the stellar mass density and surface brightness in the rest-frame \textit{U} and \textit{V}. However, Voronoi binning effectively degrades the angular resolution of these maps. In order to recover the resolution of 0.3", light variations within the deconvolved images are used to spatially calibrate the stellar masses and surface brightness. In essence, the deconvolved images in the $K_s$ band are used to scale the stellar mass density maps as this is our best filter for probing stellar masses at our redshift range. Similarly, the surface brightness (in the rest-frame $U$ and $V$) are scaled with the respective deconvolved images that probe the rest-frame UV and optical.

To this end, the deconvolved images are first scaled so that the minimum and maximum values within each deconvolved image are between 0 to 1. An image, $x$, is therefore scaled using the following equation,

\begin{equation}
    x'_i = \frac{x_i-\mathrm{min}(x)}{\mathrm{max}(x)-\mathrm{min}(x)}
\end{equation}
Using the minimum and maximum values of the whole image (as opposed to a Voronoi bin) ensures that the relative brightness within the galaxy remains preserved. Given a measured quantity ($Q$) of a Voronoi bin, the resolution of 0.3" is recovered by scaling $Q$ using the following relation, 

\begin{equation}
    Q_i' = \frac{Q x_i'}{\Sigma x_i'}
\end{equation}
In the notation above, $i$ represents pixel elements within the given Voronoi bin. $Q x_i'$ is normalized by the sum of $x_i'$ within the bin to ensure that the inferred stellar mass or luminosities from SED fitting is preserved after rescaling.

\section{Identifying and Classifying Clumpy Structures} \label{sec:method_identify}

In the following sections, we address the spatial variations within the stellar mass density and surface brightness maps. This is done by creating normalized radial profiles \citep{Wuyts2012}. A normalized profile is a two-dimensional profile that simultaneously quantifies the colors and radial dependencies on stellar properties, and enables us to detect any spatial variations within the stellar population maps. We briefly discuss how a 2-dimensional profile is made, and the classification scheme used to identify clumpy regions. We refer the reader to \cite{Wuyts2012} for further details.

\subsection{Normalized Light and Stellar Mass Profiles}
The stellar mass-weighted center of the galaxy is adopted as a reference point for all measurements. This is motivated by the fact that the mass-weighted center should correspond to a location within a galaxy where the gravitational potential is strongest. The normalized brightness profiles are then constructed as follows. 

We first associate the morphology of each galaxy with two elliptical parameters: the position angle of the galaxy and its axial ratio. The former is obtained by measuring the position angle of each pixel element within the segmentation map (as obtained from Section \ref{sec:segmap}), and constructing a distribution of the position angles. The mode of that distribution is taken to be the position angle of the galaxy. Similarly, a distribution of the pixel elements' distance to the mass-weighted center is made, and the axial ratio of the galaxy is defined to be the ratio between the mode of the distribution and the extent of the distribution. For the rest-frame $U$ and $V$ surface brightness map, we then define the half-light radius ($R_e$) of a  galaxy to be the semi-major axis length at which an elliptical aperture contains 50\% of the respective light. This radius is derived by constructing a curve of growth using elliptical apertures (defined by the position angle and axial ratio), which are placed at the mass-weighted center. Having established a half-light radius, we measure the average surface brightness within the half-light radius ($\Sigma_e$). The surface brightness of each pixel element is then normalized by the average surface brightness, and its galactocentric distance is normalized by the half-light radius. In essence, the whole process normalizes the surface brightness of each pixel element and remaps them into a new parameter space. The constructed normalized light profile will typically decrease as a function of radius, with the exceptions being for when pixel elements have an enhanced surface brightness. These enhanced surface brightness are most likely due to star-forming clumps in high-$z$ galaxies (e.g., \citealt{Wuyts2012}).  

The same procedure is used to create the normalized stellar mass profile for each galaxy. The exception is that we compute the half-mass radius as the semi-major axis length at which 50\% of the total stellar mass of the galaxy is contained. Note that the elliptical parameters are derived based on the segmentation map (as opposed to using the surface brightness) of the galaxy, so that the shape of the elliptical aperture is driven by the outer isophotes of a galaxy, rather than by individual bright clumps.


\begin{figure*}[t!]
\centering
\subfigure[]{\includegraphics[width=0.41\linewidth]{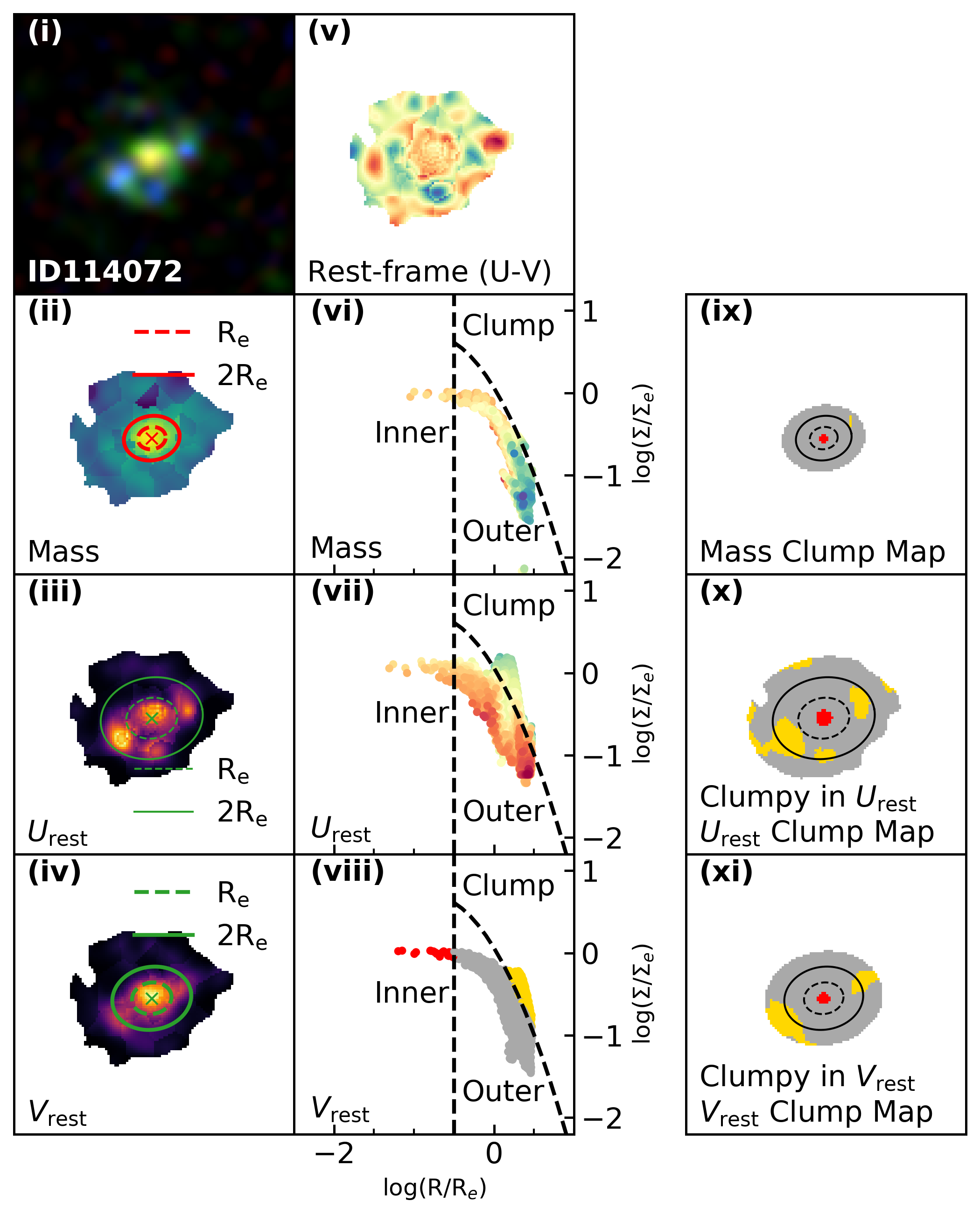}}
\quad
\subfigure[]{\includegraphics[width=0.41\linewidth]{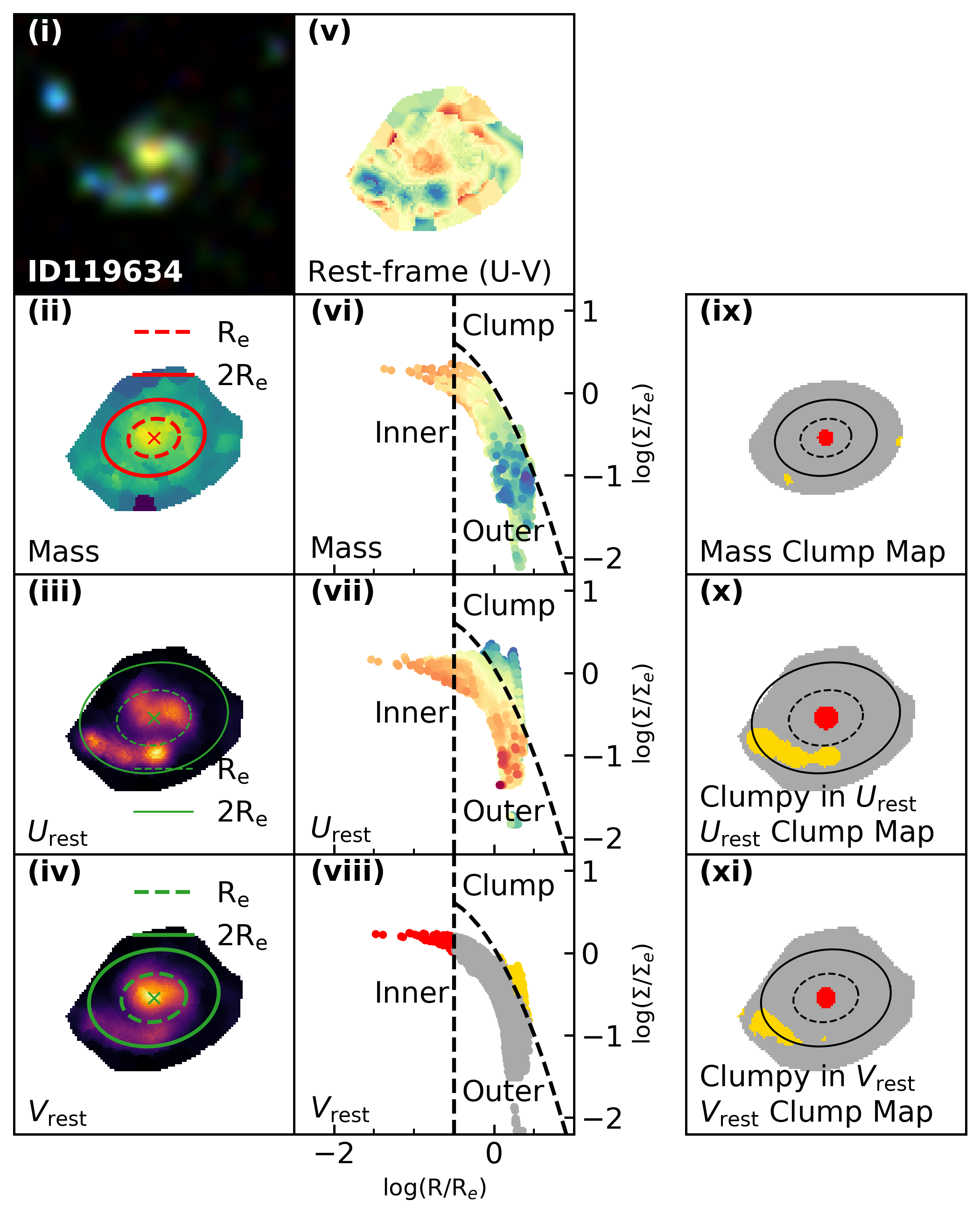}}

\subfigure[]{\includegraphics[width=0.41\linewidth]{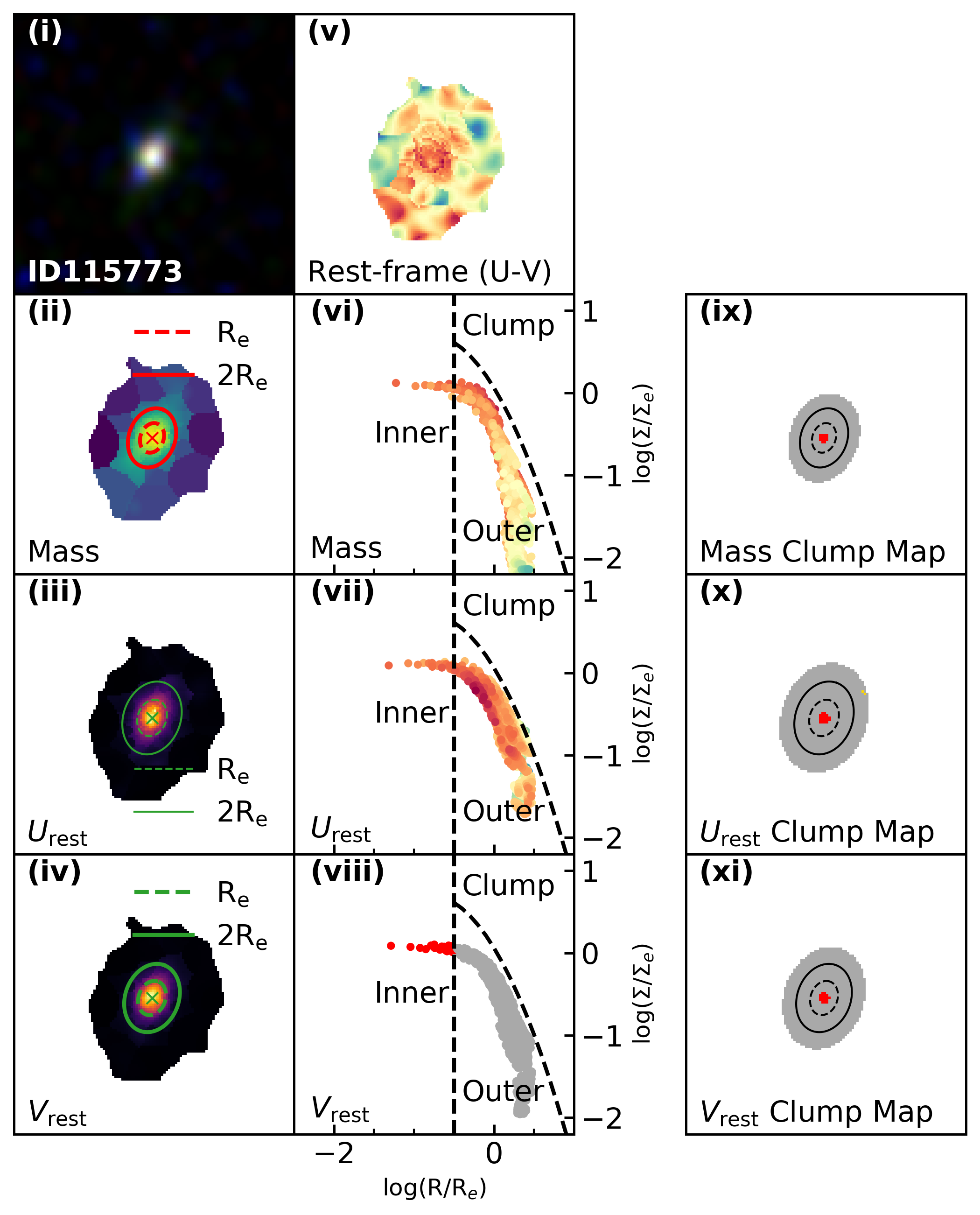}}
\quad
\subfigure[]{\includegraphics[width=0.41\linewidth]{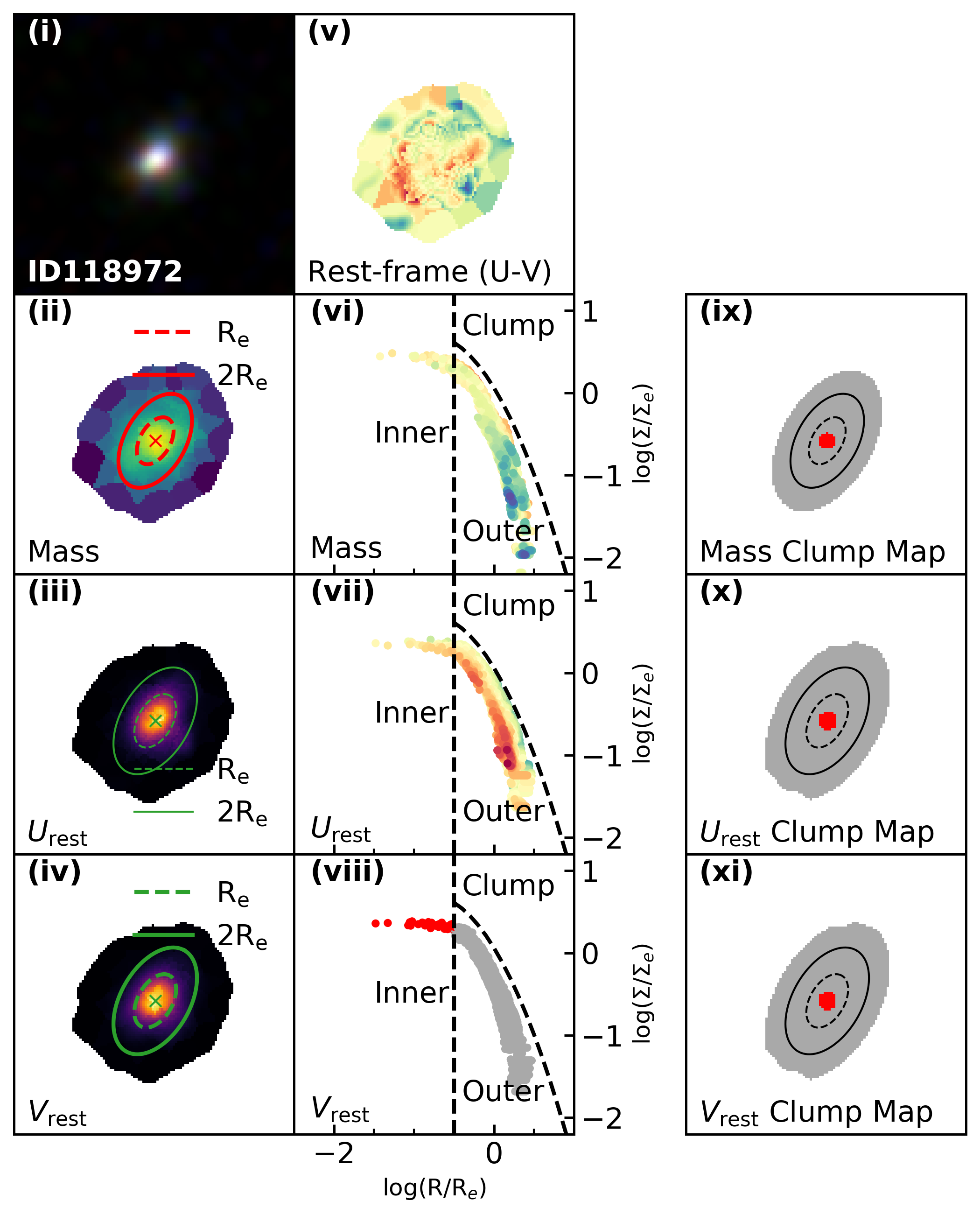}}

\caption{Examples of the normalized mass and light profiles. In each figure, panels (i), (ii), (iii) and (iv) show the composite image, the stellar mass density map, and the $U_\mathrm{rest}$ and $V_\mathrm{rest}$ surface brightness maps, respectively. The normalized mass and light profiles are shown in panels (vi), (vii) and (viii). The ($\Sigma/\Sigma_e$, $R/R_e$) space is separated into 3 different regions, where clumpy pixels are situated above the curved-dash line and represent pixel elements that have enhanced surface brightness relative to the underlying disk. We color-code pixel elements within the normalized mass/light profiles to construct clump maps as shown in panels (ix), (x), (xi), where the inner, outer and clump pixels are red, grey and yellow, respectively. (a) and (b) are examples of clumpy galaxies, while (c) and (d) are example of non-clumpy galaxies. When clumps are identified using the normalized light profiles (see panel vii and viii), the identified regions in the clump maps (panel x and xi) correspond to UV-bright clumps that are observed within the $U_\mathrm{rest}$ surface brightness and rest-frame ($U-V$) color map. }
\label{fig:clumpygalaxy}
\end{figure*}


\subsubsection{Examples of Normalized Mass/Light Profiles}

Figure \ref{fig:clumpygalaxy} shows four examples of the normalized radial profiles alongside the resolved stellar mass distribution (panel ii), $U_\mathrm{rest}$ (panel iii) and $V_\mathrm{rest}$ (panel iv) surface brightness maps. The composite image of the galaxy is also shown in panel (i), while the rest-frame ($U-V$) color map is shown in panel (v). Since clumps are regions of enhanced surface brightness compared to the underlying disk, the radial profiles can be used to identify clumpy features within the galaxies. Panels (vi), (vii) and (viii) indicate whether pixel elements are identified as clumps, or part of the inner/outer "smooth" regions based on the stellar mass distribution and surface brightness maps. The classification scheme for whether pixel elements are clumps is later discussed in details in Section \ref{sec:regionclass}.

A number of observations can be noted from the normalized profiles. All three radial profiles (panels vi, vii, viii) are flat within the half-mass/light radius. Since galaxies are typically described as Sersic profiles, the flattening of the radial profiles within the central regions suggests that our profiles are limited by angular resolution (and indeed they are, as we only deconvolve to a resolution 0.3", which is 2.4 kpc at $z \sim 1$), and implies that it may be difficult to identify centrally-located clumps. However, the concerns for the non-detection of clumps near the central regions are not severe, as previous studies of clumpy galaxies based on \textit{HST} images demonstrate that clumps are typically found near the outskirts of galaxies (e.g., \citealt{Zanella2019}). Even at higher resolution, it may still be difficult to identify clumps within the half-mass/light radius as the central brightness of galaxies are much higher. Similarly, \cite{Wuyts2012} also observed the flattening of the radial profiles with \textit{HST} images, presumably from the finite resolution of 0.18" in the F160W filter.

At larger radii within the normalized light profiles, we find that the surface brightness drops and shows a considerable scatter of ${\sim}1$ dex at $R \simeq R_e$. The large scatter in the $U_\mathrm{rest}$ (and in some cases, $V_\mathrm{rest}$) profile is driven by localized regions of enhanced brightness (e.g., ID114072 and ID119634). For the remainder of the paper, we refer to any off-center regions with an enhanced level of surface brightness as \enquote{clumps}. This nomenclature is inspired by the morphological appearance of the galaxies as shown in Figure \ref{fig:clumpygalaxy}. However, we stress that this definition encompasses all regions that have an elevated surface brightness, regardless of their geometric shapes (e.g., structures such as spiral arms). It is not intuitive whether a distinction can be made between actual star-forming clumps and other structures based on the excess of surface brightness alone. The normalized profiles also indicate that there is a color gradient; we typically find redder colors within the inner region of the objects ($R< 0.32 R_e$) as shown by the colormap of the radial profiles. The comparison between the mass and light profiles also shows that the stellar mass distribution of galaxies are generally smooth, even though their light profiles exhibit clumpy morphologies. Clumps are also easily detected within the $U_\mathrm{rest}$ map compared to $V_\mathrm{rest}$. This is expected if we consider the M/L effect where younger stellar populations are brighter at the same stellar mass. Such effect is also observed through the fraction of clumpy galaxies (see Section \ref{subsec:fclumpy_evolution}). 

\subsection{Classification Scheme for Clumpy Structures}\label{sec:regionclass}

The methodologies for detecting clumps vary from studies to studies, and no universal way have been defined. A number of methods have been used to detect clumpy structures, but the majority are based on identifying pixels with enhanced intensity through a computer algorithm (e.g., \citealt{Wuyts2012, Guo2015, Zanella2019, Huertas2020}). Our normalized profiles naturally allow us to perform a similar analysis. Following the classification scheme of \cite{Wuyts2012}, we split the normalized parameter space into three regions;
\begin{equation}\label{eqn:pixelclass}
    \begin{split}
        [\mathrm{inner}] ~~x &<-0.5 \\
        [\mathrm{outer}] ~~x &>-0.5, ~y <0.06-1.16x-x^2 \\
        [\mathrm{clump}] ~~x &>-0.5, ~y >0.06-1.16x-x^2 \\
    \end{split}
\end{equation}
where $x\equiv\log(R/R_e)$ and $y\equiv\log(\Sigma/\Sigma_e)$. Pixel elements are considered to lie in the inner, outer, or clump regime based on their location in the ($\Sigma/\Sigma_e$, $R/R_e$) parameter space. 

Note that the dashed curve of Equation \ref{eqn:pixelclass} does not truly define the light curve for all galaxies. This is why there are gaps between the normalized profiles and the dashed curve in Figure \ref{fig:clumpygalaxy}, and similar observations is also observed in the normalized profiles shown by \cite{Wuyts2012}. These dividing lines were derived by \cite{Wuyts2012} based on their stacked profiles, and are fixed for all galaxies. We argue that the same lines are also sufficient for classifying clumpy structures in our galaxies as our mass/light profiles are obtained in a similar way. Each mass/light profile is obtained by normalizing the mass density/surface brightness of each Voronoi bin by the mean mass density/surface brightness of the galaxy. While the dividing lines do not vary from galaxies to galaxies, the normalization ensures that all clumpy pixels from different galaxies are mapped to a similar region within the normalized parameter space. As shown by \cite{Wuyts2012} and this study, it is visible that these divisions are sufficient for identifying. This is illustrated in the right column of the panels in Figure \ref{fig:clumpygalaxy}, where pixels are color-coded based on their pixel type (i.e., inner, outer and clump as red, grey and yellow, respectively). In general, pixels that are identified as clumpy coincide with regions of blue color as observed in the composite images. The normalized radial profiles also illustrate this, as clumpy pixels generally have blue rest-frame $(U-V)$ color. Furthermore, for cases where the galaxies' light profile (e.g., bottom panels ofFigure \ref{fig:clumpygalaxy}) are smooth no clumps are detected.  

However, the distinction between a clumpy and non-clumpy galaxy is still subjective: are all galaxies with clumpy pixels \enquote{clumpy}? In general, studies of star-forming clumps typically use the ratio of the clumps' luminosity to the total luminosity of the host galaxy (i.e., the fractional luminosity) to make the distinction. \cite{Wuyts2012} defined clumpy galaxies to be galaxies with clumpy pixels that contribute more than 5\% of the total luminosity. On the other hand, \cite{Guo2015} and \cite{Shibuya2016} defined clumpy galaxies as those that have off-center clumps with a fractional luminosity of at least 8\%. This threshold was derived by comparing the fractional luminosity of high-$z$ star-forming clumps and local star-forming regions that were artificially redshifted. The fractional luminosity of 8\% represents a point where the number counts of high-$z$ star-forming clumps deviates from local star-forming regions (see \citealt{Guo2015} for more details). For consistency of comparison, we use the same definition of 8\% as the dividing line between clumpy and non-clumpy galaxies. Given that deconvolution is a new approach to studying clumps, having such consistency while measuring the clumpy fractions from the \textit{HST} and deconvolved data allows us to show that deconvolution can be a powerful tool for analyzing small-scale features within galaxies from ground-based images.

In essence, the resolved stellar mass and surface brightness maps are used to identify clumpy pixels. A galaxy is defined as clumpy in a given band if pixels originating from the clump regime contribute to at least 8\% of the galaxy's total luminosity in the respective band. Similarly, galaxies are identified as clumpy in stellar mass if at least 8\% of the total stellar mass of the galaxy originates from pixels in the clumpy regime of the normalized mass profile. 

Again, it is emphasized that the nomenclature for \enquote{clump} includes all off-center structures that have an excess of surface brightness or stellar mass, regardless of their geometric shape. In particular, galaxies that appear to be edge-on (or galaxies that are highly eccentric with $e>0.8$) can have pixels that lie further along the $R/R_e$ axis in the clumpy region within the normalized profile plots. We remove these galaxies from the analysis to avoid biasing our sample of clumpy galaxies. The eccentricity of the galaxies are calculated using the segmentation maps from Section \ref{sec:segmap}, where we obtain the semi-major and semi-minor axes by fitting an ellipse to the edge of the segmentation maps. The total number of galaxies that are not eccentric, and therefore is included in the analysis of the paper is 20,185. For the rest of the paper, we refer to non-clumpy galaxies as \enquote{regular} galaxies.

\section{The Clumpy Fractions} \label{sec:clumpfrac}

\begin{figure*}[!ht]
\centering
\includegraphics[width=0.85\textwidth]{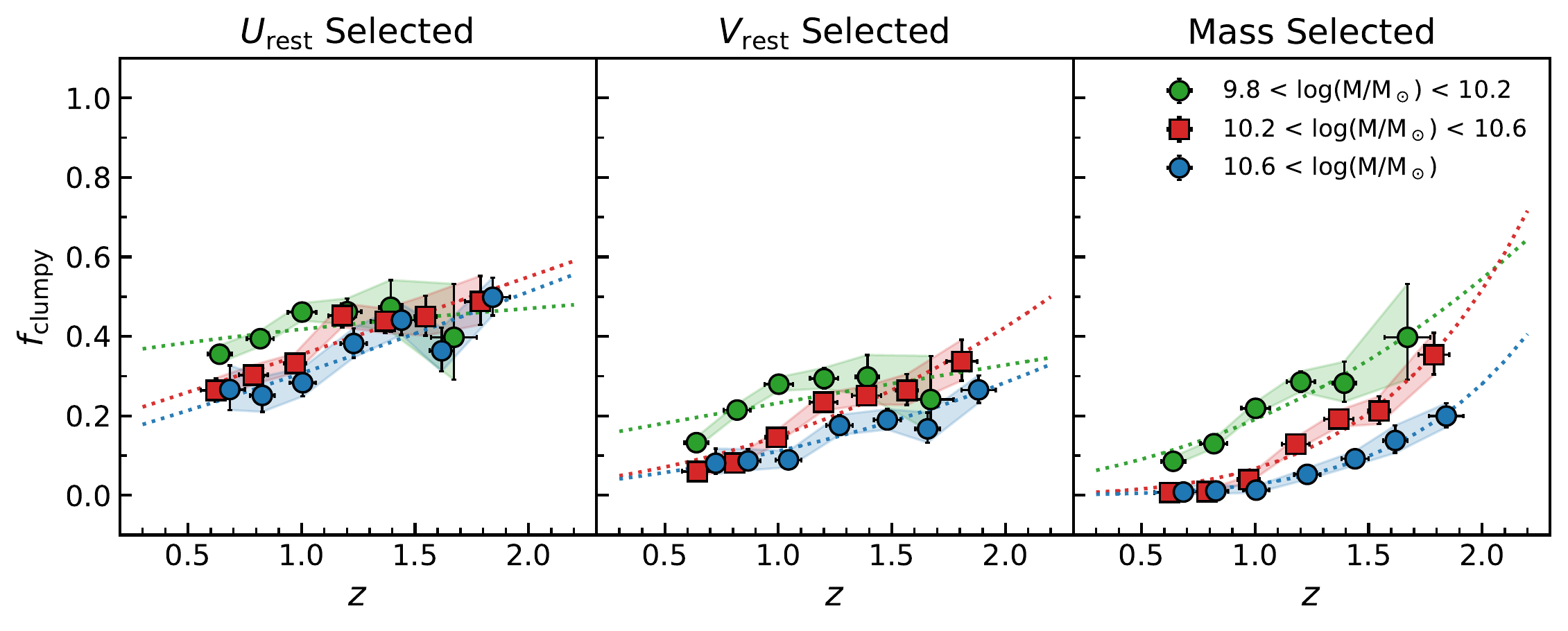}
	 \caption{The clumpy fractions for SFGs at $0.5<z<2$. From left to right, clumpy galaxies are identified based on the rest-frame \textit{U} luminosity, \textit{V} luminosity, and the stellar mass distribution map. The green, red and blue markers are the clumpy fraction of the low, intermediate, and high-mass bin respectively. The shaded region represents that errors derived from the Poisson error of the galaxy counts in those bins. The best-fit slopes for the clumpy fraction evolution are given in Table \ref{tab:cfrac_param} and are plotted as the dashed lines.}
 \label{fig:clumpy_kcorr}
\end{figure*}


In the following sections, we measure the clumpy fractions based on the surface brightness and stellar mass density of the galaxies ($f_\mathrm{clumpy}^{U\mathrm{rest}}$,  $f_\mathrm{clumpy}^{V\mathrm{rest}}$, $f_\mathrm{clumpy}^{\mathrm{mass}}$), and show the evolution of the clumpy fractions with redshift. Our measurements are also compared to other studies that are based on \textit{HST} data. Finally, we also investigate the relation between the abundance of clumps and physical properties of the host galaxy such as stellar masses and star formation rates.

\subsection{Evolution of the Clumpy Fractions} \label{subsec:fclumpy_evolution}

The evolution of the fraction of clumpy galaxies is shown in Figure \ref{fig:clumpy_kcorr}. Our galaxy sample is binned into 3 different mass bins: $9.8<\log($M$_*$/M$_\odot)<10.2$, $10.2<\log($M$_*$/M$_\odot)<10.6$ and $10.6<\log($M$_*$/M$_\odot)$. Within each mass bin, we also separate galaxies into redshift bins. The fraction of clumpy galaxies within each bin is calculated. We find that the clumpy fraction changes considerably depending on how clumps are detected, with $f_\mathrm{clumpy}$ dropping by ${\sim}$20\% when clumps are selected based on the $V_\mathrm{rest}$ surface brightness maps compared to $U_\mathrm{rest}$. When clumps are identified based on the stellar mass distribution, we find that the general trend of $f_\mathrm{clumpy}^{\mathrm{mass}}$ is similar to $f_\mathrm{clumpy}^{V\mathrm{rest}}$, but at lower amplitudes. 

The clumpy fractions are found to increase with redshifts, irrespective of the stellar mass considered. However, massive galaxies are less likely to host clumpy structures compared to low-mass systems. The slope for the evolution of the clumpy fractions also varies, where $f_\mathrm{clumpy}$ can decrease between $10-30\%$ from the highest to lowest redshift bin. The evolution of the clumpy fractions can be quantified by fitting $f_\mathrm{clumpy}$ to power-laws of the form $C(1+z)^\alpha$. The values for each fit are listed in Table \ref{tab:cfrac_param}. We find that the slope of $f_\mathrm{clumpy}^{U\mathrm{rest}}$ is shallower compared to $f_\mathrm{clumpy}^{V\mathrm{rest}}$ and $f_\mathrm{clumpy}^{\mathrm{mass}}$. The clumpy fractions for the intermediate and high-mass bins flatten out at $z<1$. However, the opposite is observed in the low-mass bin, where the clumpy fractions flatten out at $z>1$, with the exception only for clumps detected in the stellar mass maps. 

The flattening of the clumpy fractions as observed in $U_\mathrm{rest}$ and $V_\mathrm{rest}$ for low-mass systems at $z>1$ could be a S/N effect. Since galaxies are defined as clumpy based on the fractional luminosity (or mass) contribution of clumps, low S/N clumps are generally not as easily detected against the low surface brightness of low-mass galaxies. We note that while our observed clumpy fractions could be biased toward lower values for such galaxies, previous studies that used a similar definition are also affected by a similar bias. To conclude, we list the notable points of our findings;
\begin{itemize}
  \item Low-mass galaxies are generally clumpier than high-mass galaxies at all redshift, irrespective of the selection method for clumpiness (i.e., in $U_\mathrm{rest}$, $V_\mathrm{rest}$, or mass)
  \item Regardless of stellar mass, the fraction of clumpy galaxies increases toward higher redshifts. Given that S/N decreases with redshift, this trend may be even stronger than what we measure.
  \item Galaxies have higher clumpy fractions when selected in $U_\mathrm{rest}$ compared to $V_\mathrm{rest}$ or stellar mass. While the rest-frame UV trace younger stars, $V_\mathrm{rest}$ probe older stellar populations and can trace stellar mass, which can explain why $f_\mathrm{clumpy}^{V\mathrm{rest}}$ and $f_\mathrm{clumpy}^{\mathrm{mass}}$ are similar.
\end{itemize}



\begin{table}[!t]
    \centering
    \renewcommand{\arraystretch}{1.15}
    \begin{tabular}{c c c c}
    \toprule
    Mass Bin & Map & $C$ & $\alpha$ \\
    \midrule
    $10^{9.8} <$ M$_*$/M$_\odot <10^{10.2}$ & $U_\mathrm{rest}$ & 0.34 $\pm$  0.07 &  0.29 $\pm$  0.28 \\
    & $V_\mathrm{rest}$ & 0.13 $\pm$  0.06 &  0.85 $\pm$  0.57 \\
    & Mass & 0.03 $\pm$  0.01 &  2.59 $\pm$  0.38 \\
    \hline
    $10^{10.2} <$ M$_*$/M$_\odot <10^{10.6}$ & $U_\mathrm{rest}$ & 0.17 $\pm$  0.03 &  1.08 $\pm$  0.19 \\
    & $V_\mathrm{rest}$ & 0.02 $\pm$  0.01 &  2.58 $\pm$  0.39 \\
    & Mass & 0.01 $\pm$  0.01 &  5.03 $\pm$  0.64 \\
    \hline
    $10^{10.6} <$ M$_*$/M$_\odot$ & $U_\mathrm{rest}$ & 0.13 $\pm$  0.03 &  1.26 $\pm$  0.28 \\
    & $V_\mathrm{rest}$ & 0.02 $\pm$  0.01 &  2.31 $\pm$  0.44 \\
    & Mass & 0.01 $\pm$  0.01 &  5.73 $\pm$  0.44 \\
    \bottomrule
    \end{tabular}
    \caption{Listed are values obtained from fitting the clumpy fractions to $C(1+z)^\alpha$. We find that the clumpy fraction of low-mass galaxies evolves slowly with redshift compared to higher mass systems. For a given mass bin, we find that the clumpy fraction evolves strongly with redshift when using the surface mass density map, followed by the $V_\mathrm{rest}$ and $U_\mathrm{rest}$ surface brightness maps.}
    \label{tab:cfrac_param}
\end{table}

\begin{figure}[!t]
	\centering
	\includegraphics*[width=0.75\columnwidth]{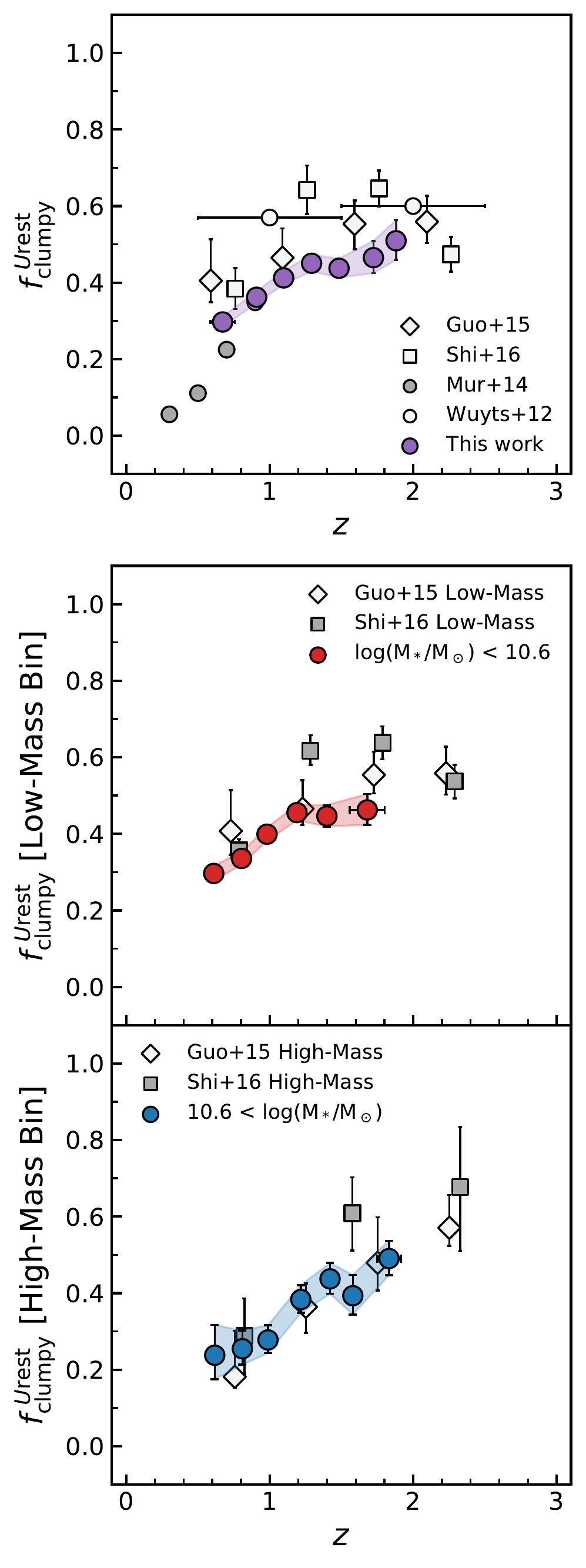}
	 \caption{Comparison of $f_\mathrm{clumpy}^{U\mathrm{rest}}$ with other \textit{HST}-based studies. The top panel shows the clumpy fractions of the whole galaxy sample, without binning galaxies by stellar masses. The middle and bottom panels shows the clumpy fractions for the low and high-mass bins, respectively. Our clumpy fractions are derived from the $U_\mathrm{rest}$ surface brightness maps. The error bars are based on Poisson counting statistic. The black and white markers are values based on \textit{HST} data, where clumps are detected within single-filter images that probe the rest-frame UV. In general, we find our values are broadly consistent with these studies, and are only slightly lower at higher redshift. }
	 \label{fig:compare_cfraction}
\end{figure}

\subsection{Comparison to \textit{HST} Studies on Clumpy Fractions} \label{subsec:fclumpy_comparison}

In Section \ref{sec:deconv_eg}, we tested our deconvolution algorithm by comparing deconvolved images to \textit{HST} images across a wide range of wavelengths and showed that deconvolved multi-wavelength images are generally consistent with their \textit{HST} counterparts. As an indirect test of how well FIREDEC can resolve structures, we also compare our measured clumpy fractions to other measurements based on \textit{HST} imaging (i.e. \citealt{Wuyts2012}, \citealt{Murata2014}, \citealt{Guo2015}, \citealt{Shibuya2016}). While our galaxy selection differs slightly from previous \textit{HST} studies (e.g., in term of stellar masses and redshift range), such a comparison still serves as a fiducial test for whether deconvolution can correctly recover spatial information within ground-based images and produce results with similar accuracy as \textit{HST}-based studies.

The top panel of Figure \ref{fig:compare_cfraction} shows the clumpy fractions for all galaxies from different studies. The clumpy fractions are derived using the rest-frame $U$-band, either by modeling the SEDs (as in the case for this study and \citealt{Wuyts2012}'s), or by using a selection of filters to probe the rest-frame UV of galaxies at different redshifts (i.e., \citealt{Guo2015, Shibuya2016}). The only exception is \cite{Murata2014}, who used the F814W filter to identify clumps in all galaxies at $0.5<z<1$. In general, the overall distribution of our $f_\mathrm{clumpy}$ is in agreement with other measured values. We find that $f_\mathrm{clumpy}$ drops from ${\sim}50\%$ to ${\sim}30\%$ between $0.5<z<2$. \cite{Wuyts2012} found a slightly evolving clumpy fraction that decreases from $60\%$ to $57\%$ between the redshift bins of $0.5<z<1.5$ and $1.5<z<2.5$. 
Similarly, both \cite{Guo2015} and \cite{Shibuya2016} reported values that are slightly higher compared to ours, but show a similar trend in the redshift evolution. Our lower fractions could be attributed to our coarser resolution of 0.3", which can blend and smear clumps. We also note that there are inconsistencies between our clumpy fractions and those reported from \cite{Murata2014}, where a steeper slope was observed in their study. While they found that $f_\mathrm{clumpy}$ drop from $35\%$ to $5\%$ between $0.2<z<1$, their measurements are based on a single \textit{HST} filter. The discrepancy could therefore be affected by the fact that galaxy morphologies change as a function of wavelength. Such effects have been observed in this study and \cite{Wuyts2012}'s, where the clumpy fractions are lower when clumps are selected at longer rest wavelengths. If we take into account of the morphological K-correction, the slope of the clumpy fractions from \cite{Murata2014} should dampen and be better aligned to ours. Considering the variety of methodologies and datasets used to distinguish star-forming clumps, the systematic differences between all studies are small.

\begin{figure*}[!t]
	\centering
	\includegraphics*[width=0.7\textwidth]{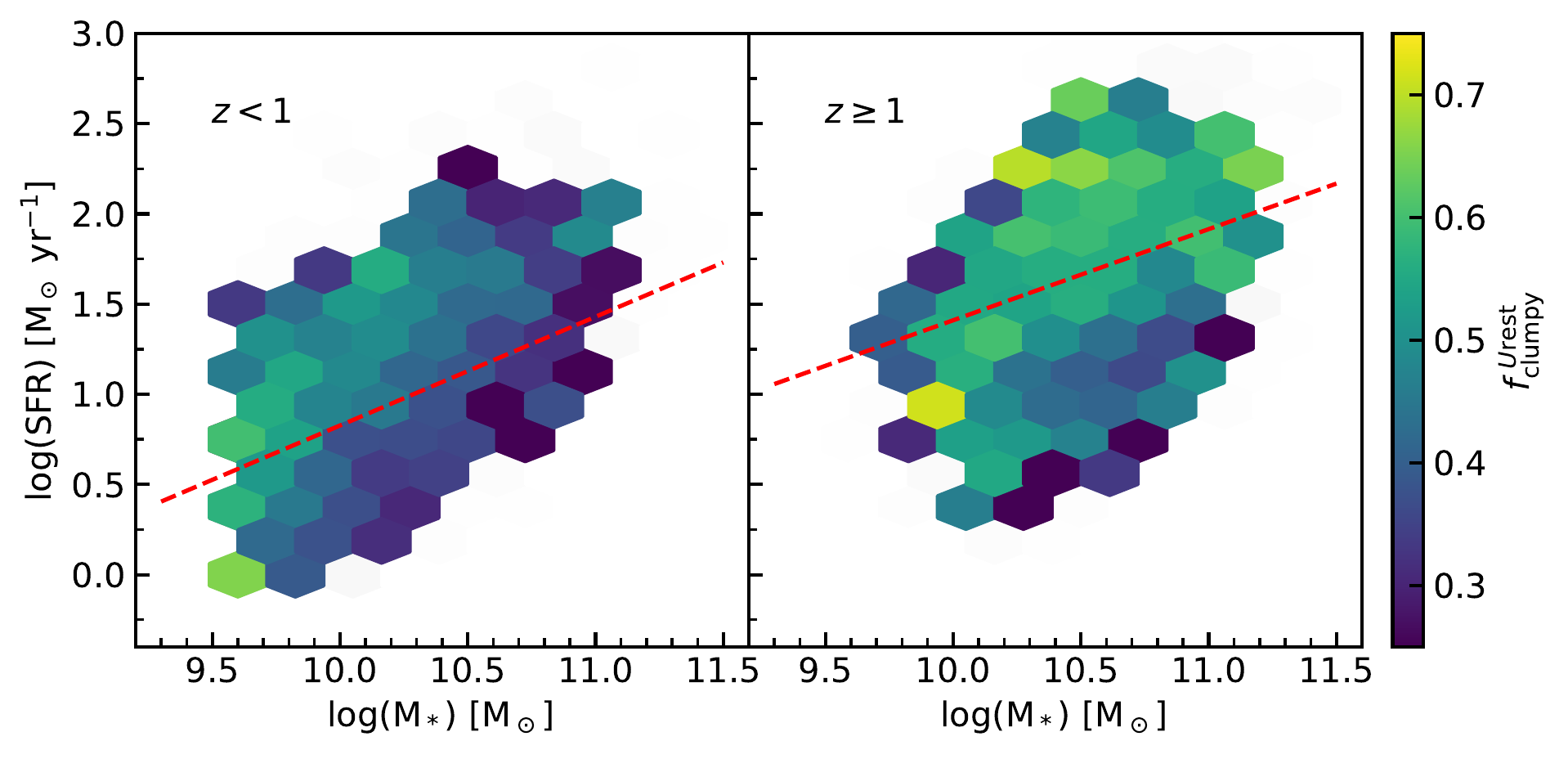}
	 \caption{Galaxies are binned within the M$_*$ and SFR parameter space, where only bins with more than 10 galaxies are shown. The colormap shows the clumpy fraction of each bin. We further separate the sample into two redshift bins, with a low-$z$ bin ($z<1$) on the left and the high-$z$ bin ($z\geq1$) on the right. The red line is the star-forming main sequence from \cite{Whitaker2012} using the fiducial redshift of 0.75 (1.5) for the low-$z$ (high-$z$) bin.}
	 \label{fig:sfms_clumpydist}
\end{figure*}

The middle and bottom panels of Figure \ref{fig:compare_cfraction} show the clumpy fraction at two different mass bins. As an approximate comparison of the literature, we split our sample into a low-mass bin at $9.8<\log($M$_*$/M$_\odot)<10.4$, and a high-mass bin at $10.4<\log$(M$_*$/M$_\odot$). In general, we find that our measurements are similar to the literature, with minor discrepancies. The evolution of $f_\mathrm{clumpy}$ is slightly different, with our values showing a slower evolution with redshift compared to both \cite{Guo2015} and \cite{Shibuya2016}. In the low-mass bin, our fraction of clumpy galaxies drops by ${\sim}15\%$ between $0.5<z<2$, compared to the ${\sim}20-30\%$ decrease observed in the \textit{HST}-based studies. At $z\sim2$, we also find that our values are lower (by $\sim15\%$) compared to those reported in \cite{Shibuya2016}. For the high-mass bin, our clumpy fraction decreases by ${\sim}30\%$ from $z\sim2$ to $z\sim1$, whereas studies such as \cite{Guo2015, Shibuya2016} reported a similar decrease, but on a shorter timescale. However, in general, our fractions are found to be mostly within the errors of other measurements.

Obtaining resolved measurements of the physical properties of clumps and their host galaxies are crucial if we truly want to understand the role that clumps play in galaxy evolution. We remind the reader that the majority of the \textit{HST}-based studies presented within this paper are limited to one filter per galaxy, and are therefore not adequate for SEDs modeling. The only exemption is the resolved studies of ${\sim}700$ galaxies at $0.5<z<2.5$ by \cite{Wuyts2012}. Within this comparison of $f_\mathrm{clumpy}$, we show that image deconvolution can be used to further extract information from existing ground-based images. Indeed, in the binary view of clumpiness, we find that the deconvolved data are largely consistent with what was observed by previous \textit{HST}-based studies. However, in addition to tracking the fraction of clumpy galaxies, the multi-wavelength deconvolved data allow us to spatially model SEDs and obtain resolved stellar mass density and surface brightness of ${\sim}$20,000 galaxies. Our larger sample also resulted in smaller statistical uncertainties in the clumpy fractions compared to the \textit{HST}-based studies.

\subsection{Dependency of Clumps on sSFRs} \label{sec:clumpdep}

Since SFRs and stellar masses are fundamental properties of galaxies, it is intriguing to question whether clumpy morphologies are associated with galaxies whose global properties deviate from those on the star-forming main sequence. In the following analyses, we further exclude galaxies with a specific star formation rate (sSFR $\equiv$ SFR/M$_*$) less than $10^{-10}$ yr$^{-1}$ in order to compare our UVJ-selected SFGs to those on the main sequence. However, our result does not change even when we include these galaxies.

Figure \ref{fig:sfms_clumpydist} shows the clumpy fraction (based on $U_\mathrm{rest}$) along the M$_*$-SFR parameter space. We bin galaxies based on their mass and SFR, and measure the clumpy fraction of each bin. Our sample is also separated into two different redshift bins; low-$z$ ($z<1$) and high-$z$ ($z\geq1$). The red dashed line denotes the star-forming main sequence relation from \cite{Whitaker2012}, using the fiducial redshift of 0.75 and 1.5 for the low and high-$z$ bin respectively. Our SFRs are taken from the COSMOS/UltraVISTA catalog, which is derived from the UV+IR luminosities. \cite{Muzzin2013} showed that the star-forming main sequence derived from the data are consistent with other measurements such as \cite{Whitaker2012}.

Galaxies have higher SFRs toward higher redshifts, which is observed as the upward shift of the star-forming main sequence relation. Interestingly, the population of galaxies that lies above the main sequence generally have a higher clumpy fraction compared to those that are situated below it. The trend persists in both redshift bins and appears to be independent of stellar mass. This is not inconsistent with Figure \ref{fig:clumpy_kcorr} if enhanced SFRs are correlated with higher clumpy fractions, such that the increase in the clumpy fractions at high-$z$ is explained as an increase in SFR within galaxies. At fixed stellar mass bins, clumpy galaxies will generally have higher SFRs compared to non-clumpy galaxies. Equivalently, clumpy galaxies have higher sSFRs compared to the average galaxies at fixed stellar mass bin. Although biases could be introduced as a result of our fine SFR and stellar mass bins in Figures \ref{fig:sfms_clumpydist}, our findings are in good agreement with the results from \cite{Murata2014}. They found that the distribution of clumpy galaxies within the M$_*$-SFR parameter space typically reside in the upper envelope of the main sequence. 

\begin{figure}[!t]
	\centering
	\includegraphics*[width=0.7\columnwidth]{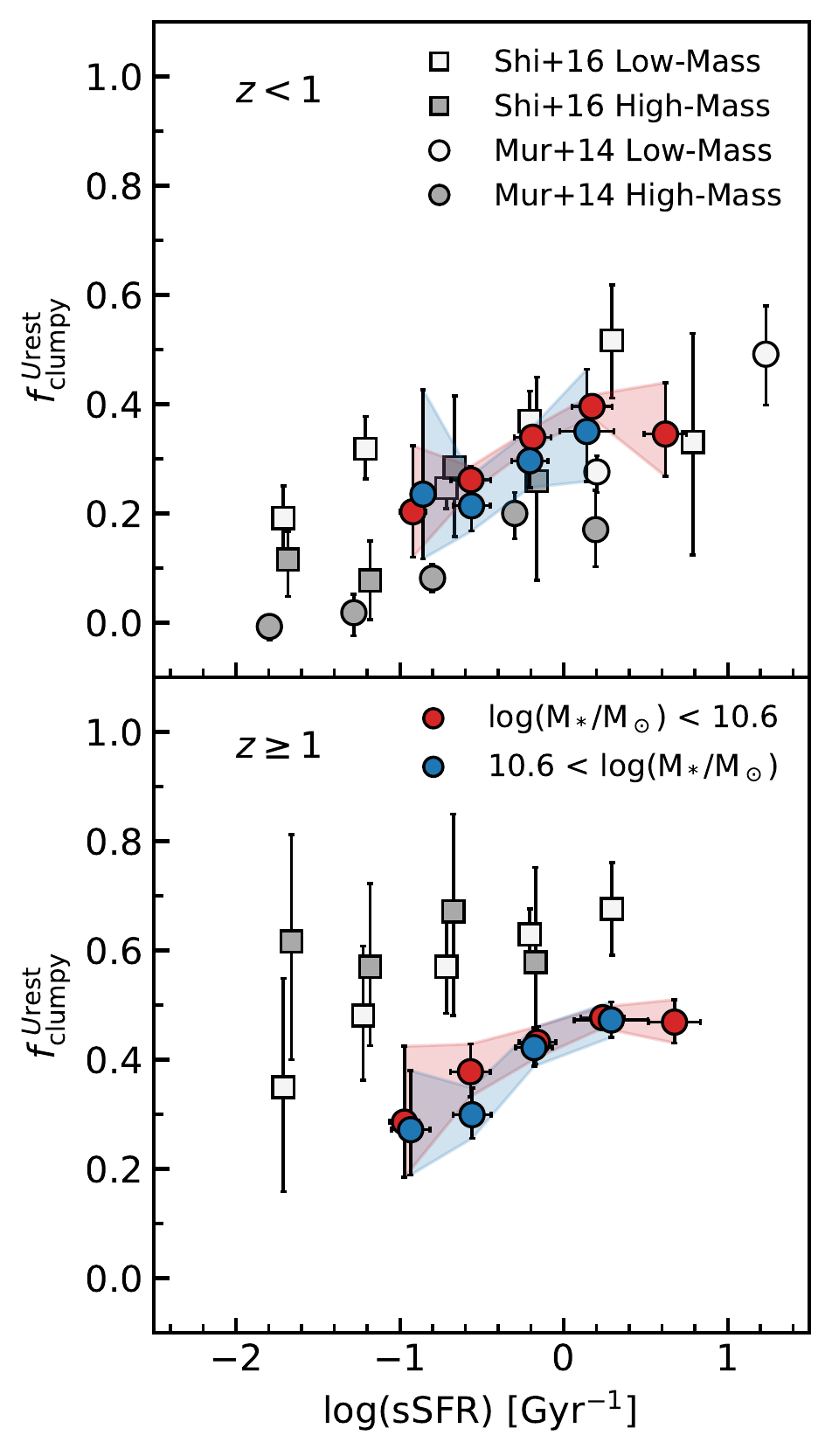}
	 \caption{Dependence of $f_\mathrm{clumpy}^{U\mathrm{rest}}$ on sSFR. The top and bottom panels show the redshift bin of $z<1$ and $z\geq1$, respectively. The blue and red markers show our measurements for galaxies with $\log(\mathrm{M}_*/\mathrm{M}_\odot < 10.6$ and $\log(\mathrm{M}_*/\mathrm{M}_\odot > 10.6$, respectively. The grey and white markers represent values from \cite{Murata2014} (circle) and \cite{Shibuya2016} (square) for similar redshift and stellar mass bins.  }
	 \label{fig:fclump_vs_ssfr}
\end{figure}

Figure \ref{fig:fclump_vs_ssfr} shows the dependence of the clumpy fraction on the sSFR, binned by stellar masses. We again separate the sample into two different redshift bins, and compare our results to similar works that are based on \textit{HST} imaging. For both redshift bins, we find that $f_\mathrm{clumpy}$ increases by ${\sim}20\%$ from our lowest sSFR bin to the highest bin. Galaxies in the high-mass bin ($\log(\mathrm{M}_*/\mathrm{M}_\odot) > 10.6$) typically have clumpy fractions that are slightly lower compared to those with lower stellar masses, although both measurements are within errors of one another. While studies such as \cite{Murata2014} and \cite{Shibuya2016} used slightly different stellar mass bins, they reported similar trends. At $z<1$, \cite{Murata2014} reported slightly lower values for the clumpy fraction as it increases from $0\%$ to ${\sim}20\%$ with sSFRs. At $z\geq1$, \cite{Shibuya2016} reported $f_\mathrm{clumpy}$ that are higher compared to ours, and their strong $f_\mathrm{clumpy}$ evolution with redshift was also observed in the bottom panels of Figure \ref{fig:compare_cfraction}. \emph{While there are differences in the normalization factor of $f_\mathrm{clumpy}$, they also found a similar trend where $f_\mathrm{clumpy}$ is increasing with sSFR in the $z\geq1$ bin, but only at the low-mass end of galaxies ($\log($M$_*/$M$_\sun)<10.6$)}. These differences could be systematic errors as various methodologies are used for detecting clumps. For example, \cite{Shibuya2016} relied on the F606W filter to detect clumps for all galaxies at $z<2$, as opposed to this work where a selection of filters is used. Their measurements are also associated with higher uncertainties, with error bars spanning between $20{-}40\%$ in $f_\mathrm{clumpy}$, compared to our errors, which benefited from having a larger sample size. On the other hand, it is possible that lower clumpy fractions are observed in deconvolved images for clumps at $z>1.5$ due to having lower S/N. 

\begin{figure}[!t]
	\centering
	\includegraphics*[width=0.7\columnwidth]{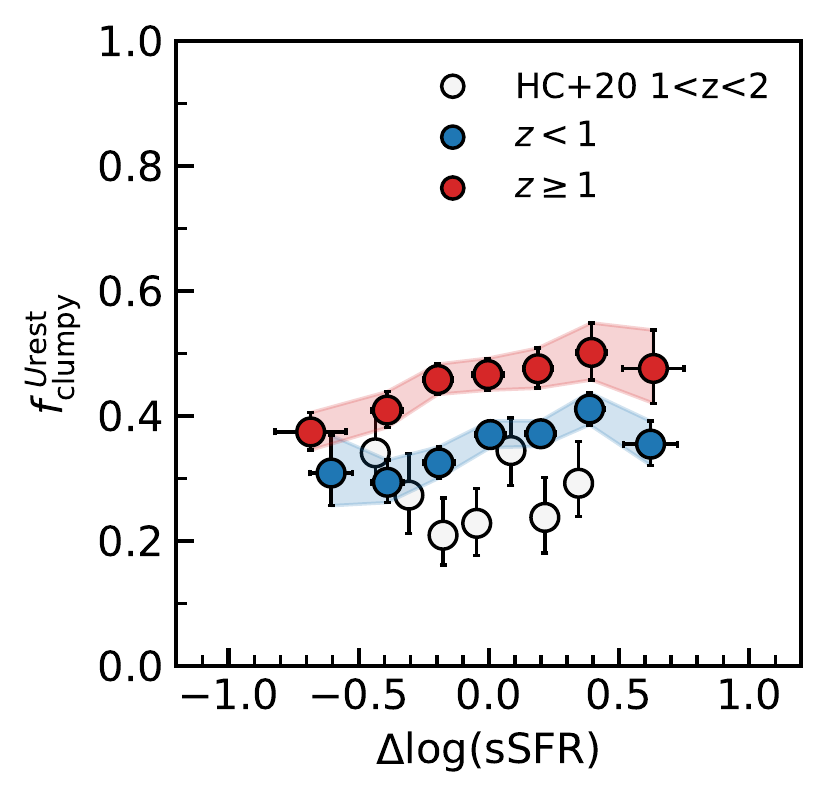}
	 \caption{The clumpy fraction in relation to the deviation of the sSFR to the mean M$_*$-sSFR relation from \cite{Whitaker2012}. Galaxies are separated into two redshift bins, with the blue and red color denoting the low-$z$ bin ($z<1$) and the high-$z$ bin ($z\geq1$), respectively. We find an increase in the clumpy fraction toward higher $\Delta\log$(sSFR). At fixed $\Delta\log$(sSFR) bin, the measured clumpy fraction is larger at higher redshift ($z\geq1$). The error bar represents the Poisson count statistic. The white markers are measurements made by \cite{Huertas2020}. They classified clumpy galaxies as those having clumps with stellar mass greater than $10^{10} \mathrm{M_\odot}$, and measured $\Delta\log$(sSFR) in relation to the M$_*$-sSFR relation from \cite{Fang2018} }
	 \label{fig:delta_ssfr}
\end{figure}

In order to investigate how the relation between $f_\mathrm{clumpy}$ and sSFR changes compared to galaxies on the main sequence, we plot the clumpy fraction as a function of $\Delta\log$(sSFR) in Figure \ref{fig:delta_ssfr}. Here, $\Delta\log$(sSFR) measures how far the specific star formation rate of our galaxies deviate from the M$_*$-SFR relation of \cite{Whitaker2012} at fixed stellar mass. Our sample is again binned into two redshift bins, and $\Delta\log$(sSFR) is binned between $-1$ to $1$. We find that the clumpy fraction is slightly dependent on $\Delta\log$(sSFR) for both redshift bins. Lower clumpy fractions are found in galaxies that have low sSFRs compared to the main sequence, with the clumpy fractions increasing by ${\sim}10\%$ at the highest $\Delta\log$(sSFR) bin. This is different what was reported by \cite{Huertas2020}, where they observed lower clumpy fractions that show no dependencies on $\Delta\log$(sSFR).

One would expect that the majority of star-forming clumps to be found in galaxies above the main sequence if \textit{in-situ} clump formation is mainly dependent on gas mass density. Since the sSFR is an indicator of the on-going star formation and could be correlated to the gas density, our observations provide a self-consistent picture that links high SFR and high gas fraction to clumpy morphologies. Indeed, the observed increase of the clumpy fractions toward high redshift in Figure \ref{fig:clumpy_kcorr} can be attributed to the increase in star formation rate within galaxies and higher gas mass fractions. We also note that our definition of clumpiness is binary. While the threshold for determining whether a galaxy is clumpy or not is physically motivated, such binary classification can affect the relation between clumps and sSFRs. In Appendix \ref{sec:c_uv_vs_ssfr}, we show that similar results to Figures \ref{fig:sfms_clumpydist} and \ref{fig:delta_ssfr} are obtained when using the fractional $U_\mathrm{rest}$ contribution of clumps to their host galaxy (hereafter, C$_{\mathrm{UV}}$, as opposed to the clumpy fraction).

\begin{figure*}[!t]
	\centering
	\includegraphics*[width=0.9\textwidth]{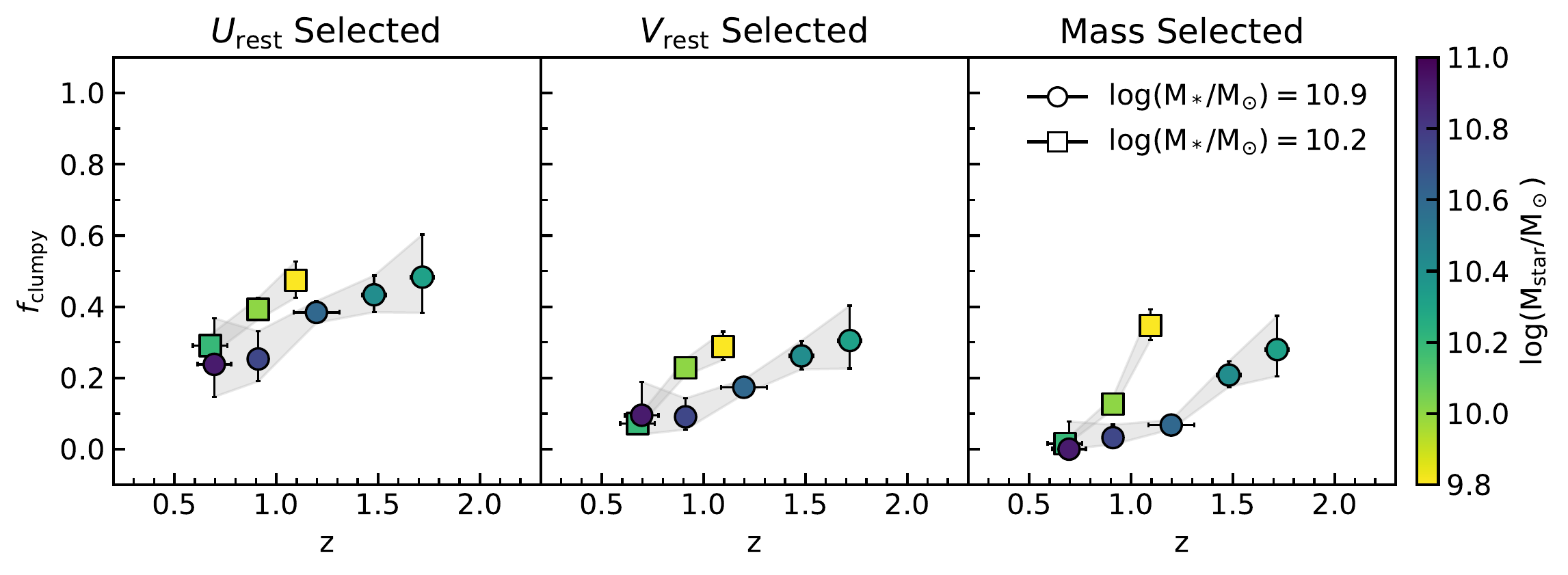}
	 \caption{The evolution of clumpy fraction for the progenitors of galaxies with masses of $10^{10.2}$ M$_\odot$ (squares points) and $10^{10.9}$ M$_\odot$ (circle points) at $z\sim0.7$. The mass evolution for progenitors is obtained from \cite{Hill2017}, by connecting the observed mass functions to the number density from \cite{Behroozi2013}. At low redshifts, clumpy morphologies are found mainly in low-mass systems, while a similar clumpy fraction is only observed in more massive galaxies at higher redshift.}
	 \label{fig:massevolv}
\end{figure*}

\section{Evolution of Clumpy SFGs Linked as Progenitors-Descendants using Abundance Matching} \label{sec:abundmatch}

The stellar mass assembled in galaxies evolves with cosmic time as stellar masses can grow from mergers and on-going star formation. Inferring the mass assembly history of galaxies can be challenging, as it requires us to accurately link progenitors to descendants. This process is complicated as galaxies can only be observed at one snapshot in time. It is possible to connect descendant galaxies to their most likely progenitors by observing how galaxy populations change in different parameter spaces (e.g., number density). The simplest method to derive the masses of progenitors is by matching their cumulative number density. This method begins with the assumption that as galaxies grow in masses, their cumulative number density would remain constant if galaxies retain their rank order in stellar mass. However, the assumption that the cumulative number density is fixed through redshift as galaxies evolve is only true if there are no mergers and galaxies have similar SFHs. The effect of mergers can be predicted using abundance matching in simulations (e.g., \citealt{Behroozi2013, Wellons2017}).


\subsection{Evolution of the Clumpy Fractions using Abundance Matching}

\cite{Hill2017} showed the progenitor mass evolution for four galaxy populations with masses of
$\log(\mathrm{M_*/M_\odot}) \sim $ 11.5, 11.0, 10.5, and 10.0 at $z = 0.1$. This was done by connecting the observed mass functions to the number densities from \cite{Behroozi2013}. We use the mass evolution calculated by those authors to trace the progenitors of two populations with with $\log($M$_*$/M$_\odot) \sim 10.9$ and $\log($M$_*$/M$_\odot) \sim 10.2$ at $z\sim0.7$. At each progenitor mass bin, we measure the fraction of clumpy galaxies. The evolution of the clumpy fractions for these progenitors is shown in Figure \ref{fig:massevolv}. We find that the clumpy fractions decrease as galaxies grow in stellar mass. This dependency suggests an interplay between clump formation and global properties such as gas availability and stellar masses. Cosmological simulations suggest that gas accretion onto massive systems is mainly through hot-mode accretion and low-mass systems through cold-mode gas accretion \citep{Keres2005}, which implies that cold gas accretion is dominant at higher redshifts when galaxies are less massive. As galaxies grow in mass, there is a suppression of gas supply, and star formation activity decreases, leading massive galaxies to passively evolve into the \enquote{red and dead} populations. However, note that cold gas accretion may also be present in massive systems at $z\sim2$ \citep{Dekel2006}, which explains how some massive galaxies at high-$z$ can exhibit clumpy morphologies. Our results therefore support \textit{in-situ} clump formation, where the unstable, gas-rich disk of high-$z$ galaxies is supplied by cold gas accretion. 

Furthermore, when comparing the evolution of the clumpy fraction for the two galaxy populations, we find that the progenitors of the low-mass population generally have higher clumpy fractions compared to the high-mass population at fixed redshift. This result illustrates the so-called \enquote{down-sizing effect} in the clumpy morphology, which have been noted by several studies (e.g. \citealt{Elmegreen2009b}, \citealt{Murata2014}): while massive clumpy galaxies exist at $z>1$, such clumpy morphologies are mainly observed in low-mass systems in the local universe. 

\subsection{Contribution of Clumps to the Mass of SFGs}

Star-forming clumps are often suggested to be an important part of the stellar mass growth of galaxies due to their ubiquity in star-forming galaxies during cosmic noon. However, measuring the contribution of clumps toward the stellar mass of the galaxy is difficult as it requires accurate measurements of the photometry of the clumps. Even with current state-of-the-art observatories, resolution effects can cause an overestimation in the size and mass of clumps (e.g., \citealt{Tamburello2017, Cava2018, Meng2020}). Noting this caveat, the following analysis looks at the fractional contribution of the stellar mass that lies within clumpy structures to the integrated mass as galaxy populations grow over time.

Following the progenitors of the two populations in Figure \ref{fig:massevolv}, we list the fractional contribution of the integrated clump mass (as identified in $U_\mathrm{rest}$) to the overall stellar mass budget of SFGs within the respective redshift bin in Table \ref{tab:clumps_contribution}. For galaxies with $\log($M$_*$/M$_\odot) \sim 10.9$ at $z\sim0.7$, we find that their clumps have a higher fractional mass contribution at earlier times: clumps account for ${\sim}5\%$ of the total mass at $z\sim1.7$, while only ${\sim}1\%$ at $z\sim0.7$. At face value, this result suggests that the stellar mass growth within galaxies relies in part on clumps. However, the mass of the progenitors are also evolving. The mean galaxy mass decreases by 0.6 dex from $z\sim0.7$ to $z\sim1.7$. This corresponds to a decrease by a factor of ${\sim}4$ in the mean galaxy mass, which is only slightly lower compared to the increase of a factor of ${\sim}5$ in the fractional mass contribution from clumps. Therefore, the total mass within clumps only increases slightly toward higher redshifts, if at all. The clumpy fraction also decreases with cosmic time (i.e., from ${\sim}50$\% to ${\sim}25$\%), suggesting that galaxies will host fewer, but more massive clumps as they evolve in mass. A similar result is obtained when we compare two populations of different masses at the lowest redshift bin ($z\sim0.7$): the more massive population have a lower clumpy fraction, but clumps from both populations have similar fractional mass contribution. When comparing two populations of similar mass, but at different redshift bins (e.g. $\log($M$_*$/M$_\odot) \sim10.2 $ at $z\sim0.7$ and $z\sim1.7$), we find that the population at high-$z$ have a higher fractional mass contribution from clumps. The clumpy fraction is also higher, supporting the scenario that clumps become more important to stellar mass growth at earlier cosmic times. 

\begin{table}[!t]
    \vspace{0.2cm}
    \renewcommand{\arraystretch}{1.15}
    \begin{tabular}{cccc}
    \toprule
    $\log$(M$_*$/M$_\odot$) & $z$ & $f_\mathrm{clumpy}^{U\mathrm{rest}}$ & $\Sigma_\mathrm{clumps}^\mathrm{mass}$/$\Sigma_\mathrm{all ~SFGs}^\mathrm{mass}$  \\
    (1) & (2) & (3) & (4) \\
    \midrule
    9.8 & 1.1 & 0.48 & 0.04 \\
    10.0 & 0.9 & 0.40 & 0.04 \\
    10.2 & 0.7 & 0.29 & 0.01 \\
    \bottomrule
    \end{tabular}
    \caption{Summary of clumps in the progenitors of galaxies with $\log($M$_*$/M$_\odot) \sim 10.2$ at $z\sim0.7$. We list the mean stellar mass of the progenitor population at different redshift bins, and present the clumpy fraction and the fractional contribution of clumps' mass to the integrated galaxy mass.}
    \label{tab:clumps_contribution}
    
    \bigskip

    \begin{tabular}{cccc}
    \toprule
    $\log$(M$_*$/M$_\odot$) & $z$ & $f_\mathrm{clumpy}^{U\mathrm{rest}}$ &  $\Sigma_\mathrm{all ~clumps}^\mathrm{mass}$/$\Sigma_\mathrm{all ~SFGs}^\mathrm{mass}$  \\
    (1) & (2) & (3) & (4) \\
    \midrule
    10.3 & 1.7 & 0.48 & 0.05  \\
    10.4 & 1.5 & 0.43 & 0.04 \\
    10.6 & 1.2 & 0.38 & 0.03 \\
    10.7 & 0.9 & 0.25 & 0.02 \\
    10.9 & 0.7 & 0.23 & 0.01 \\
    \bottomrule
    \end{tabular}
    \caption{Summary of clumps in the progenitors of galaxies with $\log($M$_*$/M$_\odot) \sim 10.9$ at $z\sim0.7$. Similar to Table \ref{tab:clumps_contribution}, we list the clumpy fraction and the fractional contribution for the progenitors. }
    \label{tab:clumps_contribution_2}
\end{table}

We again remind the reader that the nomenclature for clumps encompasses all structures that have an enhanced surface brightness (or mass) with respect to the underlying surface brightness. The measured clump mass in this Section is based on the identification of clumps using the normalized light profile, and effects such as clumps blending and background flux are not taken into account. These effects may cause an overestimation in the measured mass values. In future studies, it will be beneficial to isolate individual clumps and measure their properties (e.g., \citealt{Guo2018}). 

\section{Discussion}


At fixed stellar masses, clumpy SFGs are more likely to have higher SFRs compared to regular SFGs. The distribution of clumpy and regular SFGs within the M$_*$-SFR parameter space (Fig.~\ref{fig:sfms_clumpydist}) is not surprising and can be explained within the context of disk instability. The stability of a gas disk is described by the Toomre's $Q$ parameter \citep{Toomre:1964},

\begin{equation} \label{eqn:toomre}
    Q_\mathrm{g}\propto\frac{ v_c \sigma_\mathrm{g}}{G r \Sigma_\mathrm{g}},
\end{equation}
where $\sigma_\mathrm{g}$ and $v_c$ represent the gas velocity dispersion and circular velocity, and $\Sigma_\mathrm{g}$ is the gas density at radius $r$. The gas disk is considered to be stable if $Q>1$. It is clear that in the simple case of high cold gas density, the gas disk is more susceptible to gravitational instabilities, and can fragment into clumpy star-forming regions. The relation between gas density and the star formation rate density was established by \cite{Kennicutt:1998}, and is observed to persist for high-$z$ galaxies \citep{Genzel2010}. If we assume that this relation holds for our galaxies, then at fixed stellar masses, galaxies with high sSFRs (i.e., galaxies above the star-formation main sequence) are expected to have high gas density and subsequently low Toomre parameter values. This could explain the relation between $\Delta\log$(sSFR) and the clumpy fractions in Figure \ref{fig:delta_ssfr}. Indeed, the formation of massive clumps within a gravitationally unstable gas-rich disk have been often noted in the literature (e.g., \citealt{Escala2008, Dekel_2009}). It has also been observed in numerical simulations, where massive clumps are more common in galaxies with a high gas fraction compared to those with a lower gas fraction \citep{Fensch:2020}. 

Similar relations between the morphologies, star formation rate and gas fraction are also observed within the local universe. Locally, galaxy morphologies are based on the Hubble Sequence, with late-type galaxies referring to spiral and irregular galaxies, and early-type galaxies referring to ellipticals. \cite{Eales2017} found that morphologies change gradually with respect to sSFRs and stellar masses, where late-type galaxies typically have higher sSFRs at lower masses and early-type galaxies are more massive and have lower sSFRs. As late-type galaxies tend to have a higher cold gas reservoir compared to early-type galaxies \citep{Calette2018}, the difference in the morphology of nearby galaxies can also be explained through gas availability and star formation. 


Finally, the Toomre parameter suggests a crucial relation between the gas kinematics and the gas fraction. If the disk is expected to be marginally stable, we can set $Q_\mathrm{g} = 1$ and rearrange Equation \ref{eqn:toomre} to express $v_c/\sigma_\mathrm{g}$ in term of the gas fraction (see \citealt{Genzel2011} and \citealt{Glazebrook:2013}), 
\begin{equation} \label{eqn:gasfrac_and_kin}
    \frac{v_c}{\sigma_\mathrm{g}} \propto \frac{1}{f_\mathrm{g}}
\end{equation}
At large scales, the kinematics of high-$z$ galaxies are found to be mostly rotating disks, but with rather large gas velocity dispersion ($\sigma\approx 30{-}90$ km s$^{-1}$; see \citealt{Forster2009}). 
Theoretically, lower $v_c/\sigma_g$ values and larger gas mass fractions in high-$z$ galaxies naturally lead to the larger star-forming complexes in comparison to the less gas-rich and less turbulent disks of low-$z$ galaxies (e.g., \citealt{Escala2008}). 
What drives large velocity dispersion in high-$z$ galaxies and why such high dispersion are not observed in the local universe are both interesting questions. Galactic disks are established to be in global equilibrium, and the marginally stable state ($Q\sim1$) is suggested to be due to a self-regulating feedback loop. In the state where $Q<1$, mechanisms that drive turbulent motion are required to bring the disk back to $Q\sim1$. 

The picture of gas being accreted onto high-$z$ galaxies is a possible explanation for the initial turbulent state of the disk, while processes such as clump formation/interaction and star formation feedback are suggested to maintain turbulence (e.g., \citealt{Dekel_2009,Ceverino:2010, Krumholz:2016,Krumholz:2018}). While the relation between sSFRs and the clumpy fraction supports the picture of \textit{in-situ} clump formation, we note that these results can also be explained through \textit{ex-situ} mechanisms such as frequent gas-rich minor mergers. Such events can induce local starbursts and produce clumpy morphologies by increase the gas density locally, while decreasing $Q$ to be less than one. 



\section{Conclusion}

We use finite resolution deconvolution (FIREDEC) to resolve galaxies at the kiloparsec-scale in ground-based images of the COSMOS field. The field covers an area of ${\sim}2$ square degrees. A total of 20,185 star-forming galaxies at $0.5<z<2$ are selected to satisfy our criteria of a color, stellar mass, and S/N cut. We deconvolve galaxies to a target angular resolution of 0.3",  corresponding to a physical size of ${\sim}2.4$ kpc for the redshift range considered. Listed below are a summary of our conclusions:

\begin{itemize}
    \item We compare the multi-wavelength deconvolved images with the corresponding \textit{HST} imaging (where available) in Figures \ref{fig:multi-dec} and \ref{fig:multi-dec-2}, finding that resolved clumpy structures in deconvolved images are broadly consistent with those observed in the \textit{HST}/ACS images. Some deconvolved images may have low S/N, causing deconvolved noise look similar to clumpy structures. However, noise is not correlated across the photometric filters, so the correct clump properties can still be inferred from SEDs modeling.
    
    \item Section \ref{subsec:fclumpy_evolution} shows the relation between the fraction of clumpy star-forming galaxies, redshift and stellar mass. We find $f_\mathrm{clump}^{U_\mathrm{rest}}$ evolves with redshift, increasing from $30\%$ to $50\%$ between $z\sim0.5{-}2$ (top panel of Fig. \ref{fig:compare_cfraction}). The clumpy fractions are found to be dependent on stellar mass where massive galaxies have lower clumpy fractions, regardless of the ways clumps are detected (Fig. \ref{fig:clumpy_kcorr}). Galaxies also have higher $f_\mathrm{clump}^{U_\mathrm{rest}}$ compared to $f_\mathrm{clump}^{V_\mathrm{rest}}$ or $f_\mathrm{clump}^{\mathrm{mass}}$, while $f_\mathrm{clump}^{V_\mathrm{rest}}$ and $f_\mathrm{clump}^{\mathrm{mass}}$ are similar. This is expected since the rest-frame UV radiation come from younger stellar populations, while the rest-frame optical trace older stellar populations and stellar masses within galaxies. 
    
    \item In Section \ref{subsec:fclumpy_comparison}, we compare our clumpy fractions to the clumpy fractions derived from \textit{HST} images, finding that our measurements are consistent with those from the literature (Fig. \ref{fig:compare_cfraction}). Our measurements have smaller uncertainties due to a larger sample size. We also verify that there are no systematic differences between the fractions measured from the deconvolved and \textit{HST} images. 
    
    \item We find a larger fraction of clumpy galaxies located along the upper envelope of the star-forming main sequence as shown in Figure \ref{fig:sfms_clumpydist}. Similarly, at fixed stellar masses, clumpy galaxies have higher SFRs compared to the average star-forming galaxies. If the sSFR is correlated with gas density, this result supports the picture of \textit{in-situ} clump formation, where continuous gas accretion leads to disk instability and fragmentation. 
    
    \item We trace the progenitors of two galaxy populations at $z\sim0.7$ in Section \ref{sec:abundmatch}. Figure \ref{fig:massevolv} shows the clumpy fractions decreasing as galaxies grow in mass. This is consistent with a down-sizing picture, where clumpy morphologies are found mostly in low-mass systems in the local universe.  
    
    \item Following the progenitors of a galaxy population with $\log($M$_*/$M$_\odot) \sim10.9 $ at $z\sim0.7$, we find that the mass residing in clumps account for ${\sim}5\%$ of the total mass budget of galaxies at $z\sim1.7$. The fractional mass contribution from clumps decreases to  ${\sim}1\%$ at $z\sim0.7$. However, these galaxies are growing in mass, so the total mass within their clumps may actually remain constant through redshift. 
    
    \item When comparing two different galaxy populations of similar mass, $\log($M$_*/$M$_\odot) \sim10.2 $, but at different redshift ($z\sim0.7$ and $z\sim1.7$), we find that clumps in the high-$z$ population have a higher fractional mass contribution population.
\end{itemize}

Considering the limitations of image deconvolution, one of the notable results is that our measurements are largely consistent with those from previous \textit{HST} studies. In addition, we overcome the issues that previous studies (and future \textit{HST}/\textit{JWST} studies) have in term of limited filter coverage and sample size. The main advantage of image deconvolution is the access to a rich sample of galaxies from wide-field and multi-wavelength surveys such as COSMOS. Indeed, this study already presents that largest sample of distant galaxies with resolved imaging. It could also be very beneficial to use image deconvolution in synergy with future ground and space-based surveys (e.g., \textit{Large Synoptic Survey Telescope} and \textit{Euclid Space Telescope}). Considering the needs for resolved observations and given what deconvolution can provide, it is inviting to use deconvolution to address the topics of galaxy formation and evolution. 

\acknowledgements

We would like to thank the anonymous referee for the constructive feedback. V. Sok would also like to thank Fr\'ed\'eric Courbin for the interesting discussions on image deconvolution. The computations were performed on the Lesta computer cluster at the University of Geneva.\\

\software{EAZY \citep{Brammer2008}, FAST/IDL \citep{Kriek2009}, FAST/C++ (developed on \href{https://github.com/cschreib/fastpp/releases/tag/v1.3.1}{Github}), scikit-image \citep{Walt2014}, SExtractor \citep{Bertin1996}, vorbin \citep{Cappellari_2003}}

\appendix
\section{Determining the Lagrange Parameters} \label{append:lagrange}
Solutions to deconvolution are iteratively solved by minimizing the cost function, 
\begin{equation}
    C(B) = \sum_i \Big[\frac{D- (PSF*B)}{\sigma}\Big]^2_i + \lambda H
    \label{eqn:gencostfun}
\end{equation}
The first term is a chi-squared term that relates the solution ($B$) to the observed image ($D$) given a PSF ($P$). The second term is a regularization term weighted by a Lagrange parameter, $\lambda$. Since a similar cost function is used to numerically model the PSF of a star (see Section \ref{sub:firedec}), there are two different and important Lagrange parameters for deconvolution. For here on, we denote the Lagrange parameter as $\alpha$ when referring to the deconvolution of $D$ to obtain $B$, and $\beta$ when referring to the numerical fit of the star. The purpose of the second term in Equation \ref{eqn:gencostfun} is to mitigate noise enhancement during deconvolution. Since the regularization term contains information of high frequency components for $B$, low Lagrange parameters mean that there is minimal penalization on noise that arise during deconvolution, which can result in \enquote{noisy} deconvolved images. On the other hand, a high value of $\alpha$ and $\beta$ will result in smoother images as noise (and subsequently structures) are harshly penalized. In the following sections, we discuss how the best value for $\alpha$ and $\beta$ are determined.

\subsection{Numerical Fit}

\begin{figure*}[b!]
        \centering
        \includegraphics[width=0.95\textwidth]{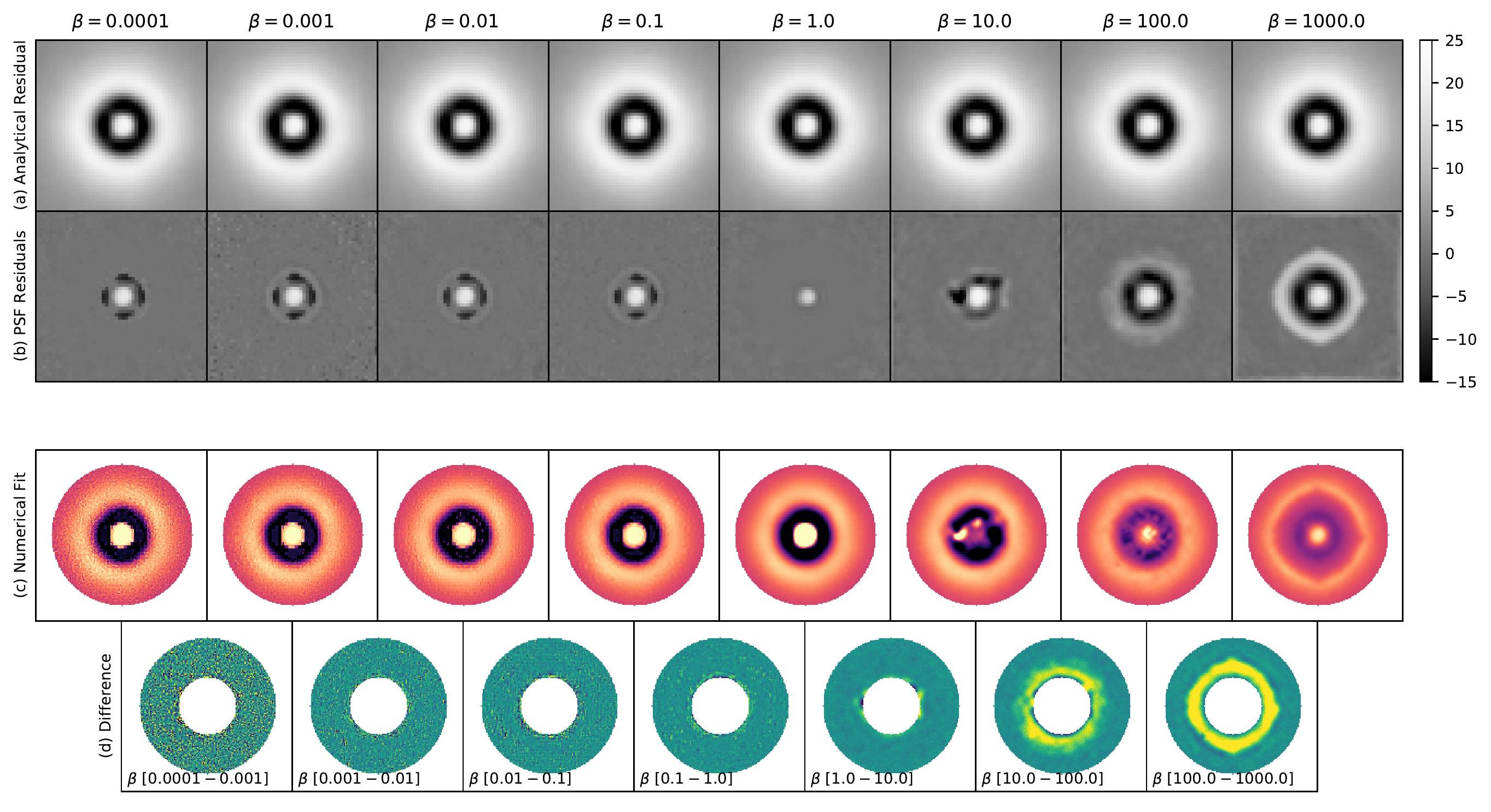}
         \caption{The numerical fit as a function of $\beta$. Row (a) shows the residual of the star after subtracting the analytic fit. The PSF wings, seen as the halo-like structure, cannot be modelled by the analytic fit alone and require a numerical fit. Row (b) shows the residual of the star after subtracting both the analytic and numerical fits. At higher $\beta$, the numerical fit fails to fully capture the PSF wing, whereas lower $\beta$ values allow for a more accurate description. Row (c) and (d) show the numerical fit and the difference between each numerical fit, respectively. While lower $\beta$ values still allows for the PSF wings to be modelled, the numerical fit is generally noisier as a result of no noise suppression in the cost function.}
         \label{fig:numfits}

        \bigskip 
         \includegraphics[width=0.95\textwidth]{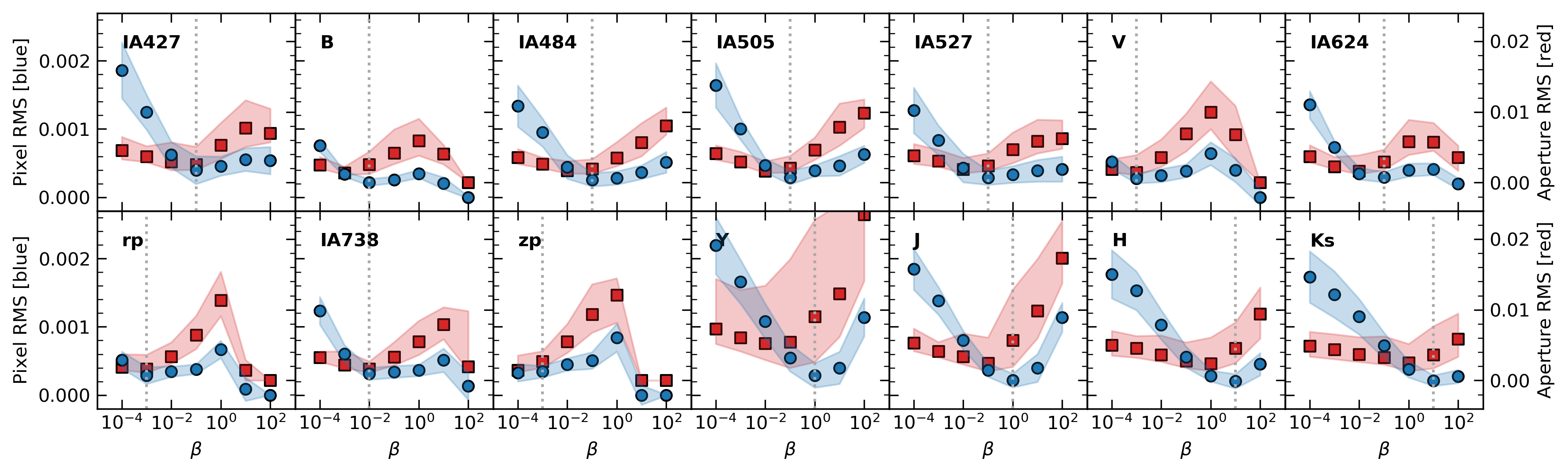}
         \caption{The RMS as a function for $\beta$ for different filter. We plot the mean pixel and aperture RMS as the blue and red points. The error is taken as the standard deviation for all the stars in COSMOS. A good $\beta$ should saddle the two RMS. The vertical dotted line indicates the value we used for each filter. }
         \label{fig:lambdnum_sum}
    \end{figure*}

Figure \ref{fig:numfits} shows an example of how different values of $\beta$ affect the numerical fit of a star observed in the $J$ filter. Row (a) shows the residual of the analytic fit. There is a halo-like structure that is not properly fitted by the analytic model. The goal of the numerical fit is capture this structure. Row (b) shows the residual of the star after subtracting both the analytic and numerical fits. We find that the halo-like structure is still present in these residual maps for higher values of $\beta$ ($\beta>100$), but disappear at lower $\beta$ values. While the residuals of row (b) imply that a low value of $\beta$ should always be used, the numerical fit itself (i.e.,  row c) is becoming noisier toward low values of $\beta$. This is shown in row (d) where we take the difference between the numerical fits. It is apparent that the pixel-to-pixel variation increases toward lower $\beta$. Therefore, only a range of $\beta$ allows the numerical fit to maximally capture the halo-like structure in the analytic residual, while adding minimal noise to the numerical fit. The best value of $\beta$ for the example in Figure \ref{fig:numfits} is $\beta=1$. 

This is quantifiable by comparing the root-mean-squared errors (RMS) within row (d) of Figure \ref{fig:numfits} as a function of $\beta$. We first calculate the pixel RMS to quantify pixel-to-pixel variations in the residual maps of row (d). The pixel RMS is taken to be the standard deviation of the histogram of pixel intensities. To quantify the large-scale structures within the residual maps, we calculate the RMS of flux within apertures, where the aperture diameter is taken to be the target resolution at 0.3". These apertures are placed randomly within the residual maps of row (d). A good $\beta$ value should saddle the two RMS. Figure \ref{fig:lambdnum_sum} shows both RMS for different filters. We mark the best $\beta$ for each filter as the vertical dotted line.

    \begin{figure*}[t!]
    \centering
    \includegraphics[width=0.95\textwidth]{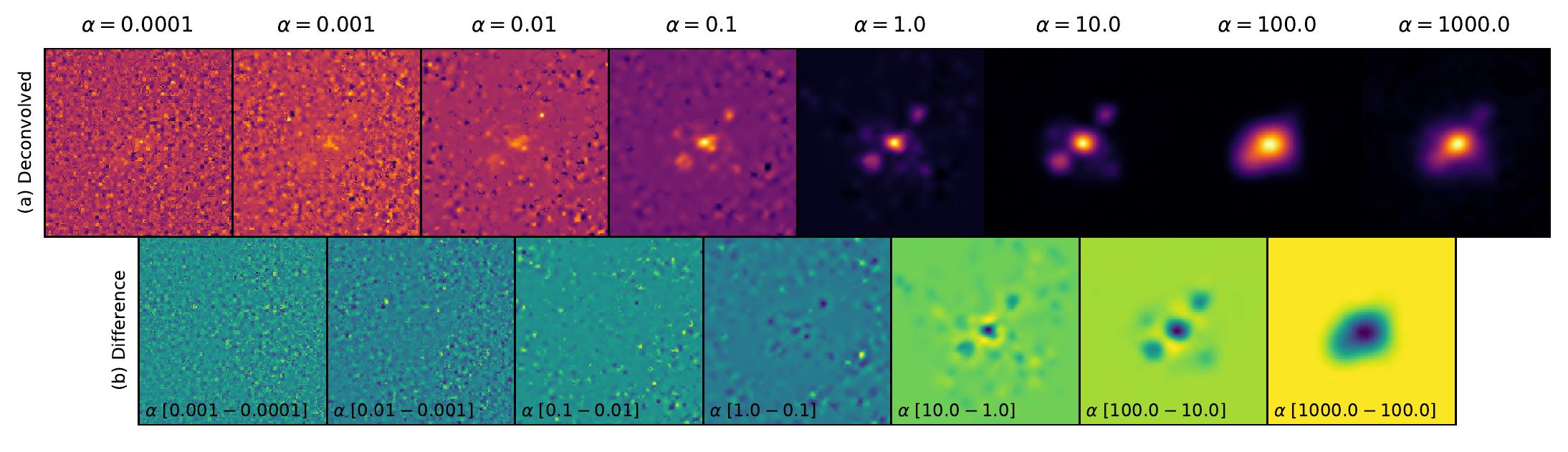}
     \caption{Deconvolved images as function of $\alpha$. Row (a) shows the deconvolved $z^+$-band image of a galaxy, with $\alpha$ increasing by a factor of ten from left to right. In general, lower $\alpha$ result in noisier images. Row (b) shows the differences between each deconvolved image. The residuals within these maps become more uniform at lower values of $\alpha$, indicating that no new information is being fitted by deconvolution. It should also be noted that the noise in the deconvolved image are enhanced as $\alpha$ is decreased. A good $\alpha$ should minimize the noise in the deconvolved image (row a), but also allow structures to be maximally fitted. The latter case is illustrated in row (b) where the residuals map become uniform between $\alpha=0.1$ and $\alpha=1$. }
     
     \label{fig:decfits}
     
     \bigskip
     \includegraphics[width=0.7\textwidth]{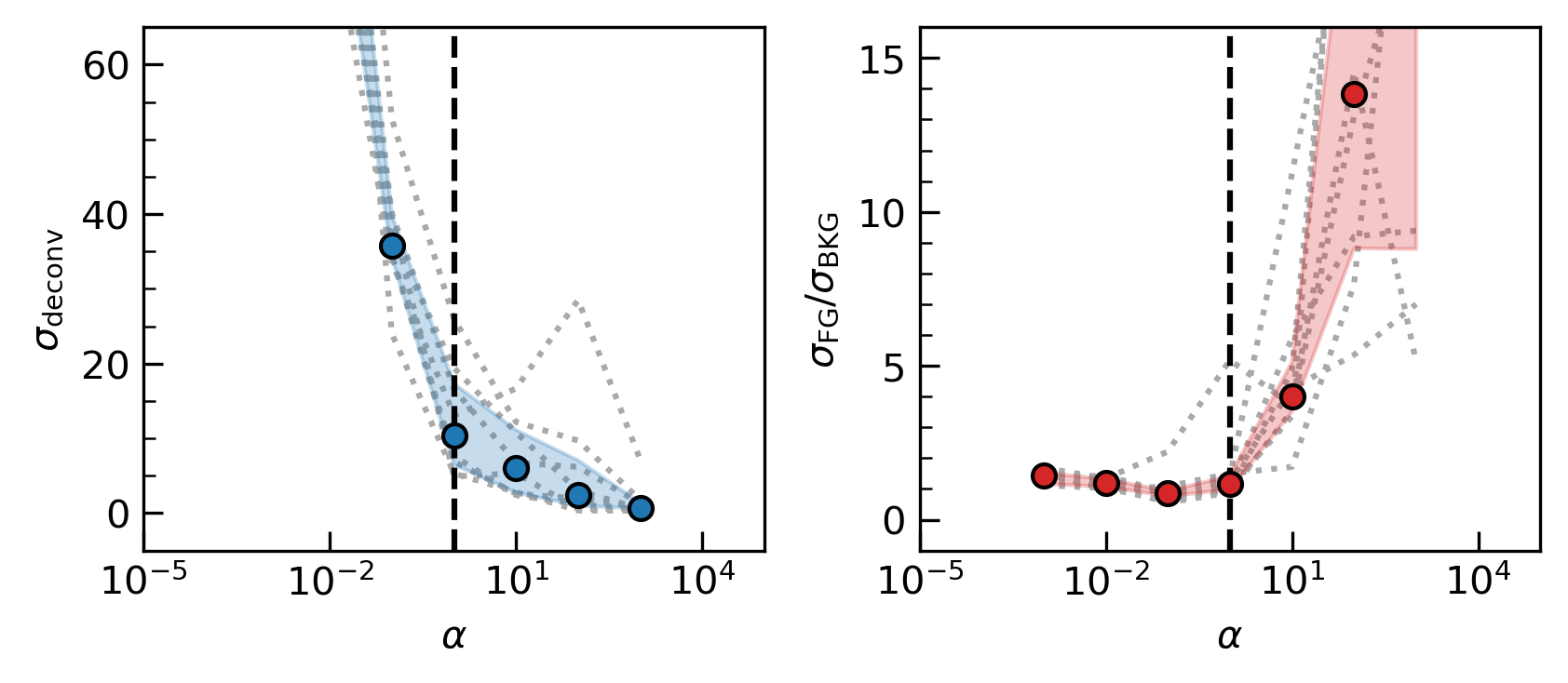}
     \caption{The left panel shows the noise for a few deconvolved $z^+$-band images as a function of $\alpha$, while the right panel shows the ratio of the aperture noise within the foreground and background of the residual maps in Figure \ref{fig:decfits}. The foreground and background of the galaxy is defined by the edge of the segmentation map for the galaxy. The grey dotted lines are values for different galaxies. The means are also plotted, where the shaded region represents the standard deviation. In general, lowering $\alpha$ allows finer structures to be fitted, but can cause noise to increase in the background. A good $\alpha$ should saddle the two curves. For the $z^+$ filter, the black dashed line at $\alpha=1$ represents a good value.}
     \label{fig:decsum}
     
     \bigskip
    \includegraphics[width=0.95\linewidth]{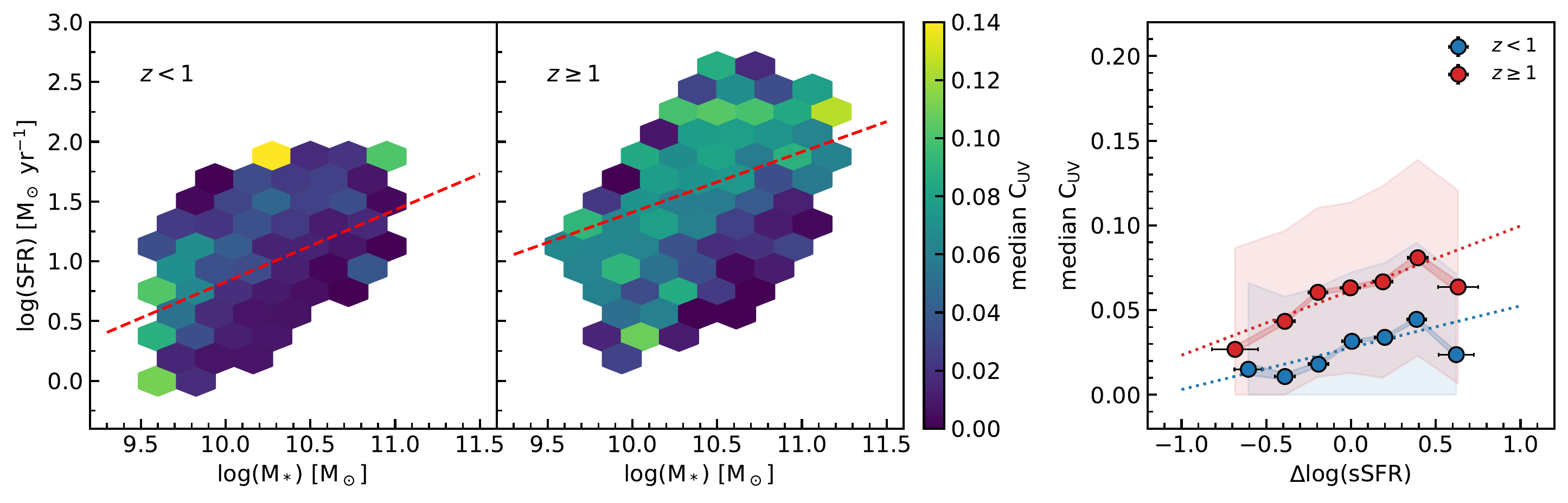}
    \caption{Similar to Figures \ref{fig:sfms_clumpydist} and \ref{fig:delta_ssfr}, but we use the median C$_\mathrm{UV}$ as opposed to the clumpy fraction. We generally find similar trends based on the clumpy fraction and C$_\mathrm{UV}$. In the left panel, the median C$_\mathrm{UV}$ is typically higher for galaxies above the main sequence. In the right panel, the median C$_\mathrm{UV}$ is observed to increase with $\Delta\log(\mathrm{sSFR})$ for both redshift bins. However, the line of best fit indicates that the slope of the relation is relatively shallow. The lighter shaded region shows the  1-$\sigma$ dispersion of C$_{\mathrm{UV}}$, while the darker shaded region shows the standard error of the median ($\sigma/\sqrt{n}$).   }
    \label{fig:delta_ssfr_Cuv}

    \end{figure*}


\subsection{Deconvolution of Galaxies}

The regularization term in Equation \ref{eqn:gencostfun} is weighted by a different Lagrange parameter ($\alpha$) when deconvolving images of galaxies. Figure \ref{fig:decfits} shows an example of how different $\alpha$ values affect deconvolution. Row (a) shows different deconvolution of the same galaxy, but with $\alpha$ increasing by a factor of ten from left to right. Row (b) shows the differences between each deconvolved image. Since the deconvolved images are obtained using the same parameters (with the exception of $\alpha$), the residual maps illustrate how $\alpha$ affects deconvolution. In particular, the deconvolved images mainly consist of noise at lower $\alpha$ ($\alpha\lesssim0.01$), whereas smoother images are obtained with higher values. The fact that the residual maps in row (b) have a uniform noise distribution demonstrates that a solution can converge by lowering $\alpha$. However, lowering the value of $\alpha$ further introduces more noise in the deconvolved image, and the solution becomes degenerate. 

We quantify the convergence by comparing the aperture noise within the foreground and background of the residual maps. The dividing line between the foreground and background of the galaxy is defined by the segmentation map of the galaxy. As the ratio between the two RMS converges to unity, it indicates that no new information are being fitted by deconvolution. On the other hand, the pixel RMS within the background of the deconvolved image is increasing as $\alpha$ decreases. This is summarized in Figure \ref{fig:decsum} for a few galaxies in the $z^+$-band. A good $\alpha$ should saddle the two curves. As shown in the example of Figure \ref{fig:decsum}, $\alpha=1$ provides a good deconvolution model. 

\section{Dependence of the Fractional Light Contribution of Clumps on the sSFR} \label{sec:c_uv_vs_ssfr}

In Section \ref{sec:clumpdep}, a slight dependency was observed between the fraction of clumpy galaxies and $\Delta\log(\mathrm{sSFR})$. However, classifying galaxies based on a luminosity cut-off (e.g., where clumpy galaxies are defined to have $\geq8\%$ of their luminosity in clumps) can blur out the relation between the abundance of clumps and $\Delta\log(\mathrm{sSFR})$. An alternative way to test this relation is to use the fractional UV contribution of clumps. Figure \ref{fig:delta_ssfr_Cuv} shows two plots similar to Figures \ref{fig:sfms_clumpydist} and \ref{fig:delta_ssfr}, but the median C$_{\mathrm{UV}}$ is used instead of the clumpy fraction. C$_{\mathrm{UV}}$ represents the fraction of UV light coming from clumps to the total UV light of the host galaxy. Comparing the left panel of Figure \ref{fig:delta_ssfr_Cuv} to Figure \ref{fig:sfms_clumpydist} shows that the clumpy fraction is correlated to the median C$_{\mathrm{UV}}$. This is not surprising as clumpy galaxies are defined to have $\mathrm{C}_{\mathrm{UV}} \geq 0.08$. Galaxies with higher C$_{\mathrm{UV}}$ are therefore generally found above the main-sequence relation.

The right panel of Figure \ref{fig:delta_ssfr_Cuv} shows the median C$_{\mathrm{UV}}$ as a function of $\Delta\log$(sSFR). The darker shaded regions show the standard error of the median, which measures the discrepancy between our sample's median and the true population's median. This is calculated as $\sigma/\sqrt{n}$, where $\sigma$ is the dispersion of C$_{\mathrm{UV}}$ measured relative to the median value and $n$ is the number of galaxies in each bin. In general, we find that the median C$_{\mathrm{UV}}$ increases with $\Delta\log$(sSFR). However, the median C$_{\mathrm{UV}}$ decreases at $\Delta\log(\mathrm{sSFR}) > 0.5$, indicating that clumps within galaxies contribute less to the total UV luminosity. A slope of $0.04\pm0.01$ and $0.02\pm0.01$ is respectively obtained for the high-$z$ and low-$z$ bin by fitting a linear line. We test the significance of the slopes using a t-test statistic, where a t-score of 4 and 2 is obtained for the high-$z$ and low-$z$ bin. These t-scores correspond to p-values that are ${<}0.05$. While this result formally suggests that a correlation exists between C$_{\mathrm{UV}}$ and $\Delta\log(\mathrm{sSFR})$, we note that the slopes are still shallow and that C$_{\mathrm{UV}}$ is highly scattered. The 1-$\sigma$ dispersion of C$_{\mathrm{UV}}$ is illustrated as the lighter shaded region.



\clearpage
\bibliography{library}{}
\bibliographystyle{aasjournal}



\end{document}